%% file: ijcai22-full.tex
\documentclass{article}
\usepackage{ijcai22}

\usepackage{times}
\usepackage{soul}
\usepackage{url}
\usepackage[hidelinks]{hyperref}
\usepackage[utf8]{inputenc}
\usepackage[small]{caption}
\usepackage{graphicx}
\usepackage{amsmath}
\usepackage{amsthm}
\usepackage{booktabs}
\usepackage{algorithm}
\usepackage{microtype}
\usepackage{booktabs} 
\usepackage[utf8]{inputenc}
\usepackage{url}
\usepackage{amsthm}
\usepackage{amssymb}
\usepackage[noend]{algpseudocode}
\usepackage{verbatim}
\usepackage{graphicx}
\usepackage[font=small,skip=0pt]{caption}
\usepackage{amssymb}
\usepackage{tikz}
\usepackage{listings}
\usepackage{booktabs}
\usepackage{array}
\usepackage{chngcntr}
\usepackage{url}
\usepackage{mathtools}
\usepackage{xcolor}
\usepackage{nomencl}
\usepackage{siunitx,xinttools,xintfrac}
\usepackage{lipsum}
\usepackage{subfig}
\usepackage{xspace}
\usepackage{enumitem}
\usepackage{wrapfig}
\usepackage{multirow}

\usepackage{pgfplots}
\usepackage{filecontents}
\usetikzlibrary{matrix}
\usepgfplotslibrary{groupplots}
\pgfplotsset{compat=newest}

\usepackage{hyperref}

\usepackage{newfloat}
\usepackage{listings}
\floatstyle{ruled}
\newfloat{listing}{tb}{lst}{}
\floatname{listing}{Listing}

\newtheorem{theorem}{Theorem}

\newtheorem{claim}{Claim}
\newtheorem{lemma}{Lemma}
\newtheorem{mydef}{Definition}

\algdef{SE}[DOWHILE]{Do}{doWhile}{\algorithmicdo}[1]{\algorithmicwhile\ #1}%
\DeclarePairedDelimiter\ceil{\lceil}{\rceil}

\makeatletter
\let\OldStatex\Statex
\renewcommand{\Statex}[1][3]{%
  \setlength\@tempdima{\algorithmicindent}%
  \OldStatex\hskip\dimexpr#1\@tempdima\relax}
\makeatother

\newcommand{\maxmp}{\textsc{MaxMP}\xspace}
\newcommand{\drand}{\textsc{Prob}\xspace}
\newcommand{\fdrand}{\textsc{FastProb}\xspace}
\newcommand{\fast}{\textsc{ThrGreedy}\xspace}
\newcommand{\gr}{\textsc{Greedy}\xspace}
\newcommand{\rgr}{\textsc{ResGreedy}\xspace}
\newcommand{\soda}{\textsc{SplitGrow}\xspace}
\newcommand{\boostapp}{Boosting Influence Spread\xspace}
\newcommand{\videoapp}{Video Summarization\xspace}

\title{Efficient Algorithms for Monotone Non-Submodular Maximization \\ with Partition Matroid Constraint}

\author{
Lan N. Nguyen
\and
My T. Thai
\affiliations
Department of Computer and Information Science and Engineering \\
University of Florida, Gainesville, Florida 32611
\emails
lan.nguyen@ufl.edu,
mythai@cise.ufl.edu
}

\begin{document}

\maketitle

\begin{abstract}
In this work, we study the problem of monotone non-submodular maximization with partition matroid constraint. Although a generalization of this problem has been studied in literature, our work focuses on leveraging properties of partition matroid constraint to (1) propose algorithms with theoretical bound and efficient query complexity; and (2) provide better analysis on theoretical performance guarantee of some existing techniques. We further investigate those algorithms' performance in two applications: \boostapp and \videoapp. Experiments show our algorithms return comparative results to the state-of-the-art algorithms while taking much fewer queries.
\end{abstract}

\section{Introduction}

Maximizing classes of set functions, generalizing submodular functions, has emerged recently due to its wide range applications in real-world problems. Among those works, non-submodular maximization subject to cardinality constraint was studied the most extensively, including but not limited to \cite{bian2017guarantees,das2011submodular,qian2018multiset,kuhnle2018fast}. 

However, cardinality constraint may not be sufficient to capture some natural requirements of various applications. For example, in many viral marketing campaigns, it is important to ensure the diversity and fairness among different ethnics and genders. These applications aim to distribute budget to feed information fairly among different groups of users while guaranteeing to maximize the influence spread in the network. Another example is data summarization. In many situations, a large data may be formed by elements of various classes. The problem, thus, aims to find a representative subset to cover the dataset's content as much as possible while imposing a constraint that the subset should  contain a number of members of each class to guarantee diversity.

Motivated by those observation, we study the following problem: Given a ground set $V$, a non-negative monotone function $f: 2^V \rightarrow \mathbb{R}^\geq$; let $V_1,...,V_k$ be a collection of disjoint subsets forming $V$ (i.e. $V = V_1 \biguplus ... \biguplus V_k$), and $b_1,...,b_k$ be $k$ integers that $1 \leq b_i \leq |V_i|$ $\forall i \in [k]$. The problem asks for:
\begin{align*}
    \max_{S \subseteq V} \{ f(S) : |S \cap V_i| \leq b_i ~\forall i \in [k] \} \tag{\maxmp}
\end{align*}
\maxmp is formally represented as monotone non-submodular maximization with partition matroid constraint. This constraint is a special case of matroid constraint and generalizes cardinality constraint. 

Non-submodular maximization beyond cardinality constraint was only received attention recently. The most recent works are \cite{chen2018weakly} and \cite{gatmiry2018non}, in which they studied the performance guarantee of \gr or \textsc{Residual} \gr (\rgr) \cite{buchbinder2014submodular} on monotone non-submodular maximization subject to matroid constraint. However, those algorithms requires $O(nK)$ queries of $f$ ($K$ is a rank of a matroid), which may not be desirable in practice. Researchers \cite{mirzasoleiman2016fast,badanidiyuru2014fast,kuhnle2018fast} have sought ways to speed up the \gr algorithm. Unfortunately, these approaches were only for cardinality constraint; or relied upon the submodularity of $f$.

To our knowledge, there exists no specific work dedicating for non-submodular maximization subject to partition matroid constraint. That leaves us open questions on: (1) With partition matroid, does there exist an algorithm with a better ratio or can we improve the ratio of the existing algorithms, whose performance guarantees have been proven with a matroid constraint? (As partition matroid is a special case of matroid constraint, perhaps we can get a tighter ratio if we only considered the partition matroid.) (2) Can we leverage partition matroid properties to devise approximation algorithms with more query-efficient? 

\textbf{Our Contribution}. In this work, we focus on answering those two above questions. First, to quantify the non-submodularity of a function, we introduce Partition Matroid Curvature $\alpha$ and Partition Matroid Diminishing-Return ratio $\gamma$. These two quantities are derived from a same concept with the diminishing-return ratio \cite{lehmann2006combinatorial,bogunovic2017robust} and generalized curvature \cite{bian2017guarantees,conforti1984submodular,iyer2013curvature} but have more relaxed requirement.

Our main contribution is to introduce a novel approximation algorithm, named \drand, with approximation ratio of $(1/\gamma^\prime -1 + \alpha^\prime)(1-1/\Theta(\max_{i\in[k]} |V_i|)) + 1$ where $\gamma^\prime$ and $\alpha^\prime$ are non-trivial and obtainable bounds of $\gamma$ and $\alpha$. \drand's novelty lies in a random process of selecting a new element, in which the algorithm introduces a new probability distribution among non-selected elements. That probability distribution is a key for \drand to obtain its ratio. Furthermore, by utilizing a sampling technique to reduce searching space, we propose \fdrand, an algorithm improving from \drand with efficient query complexity of $O(n\ln^2 \sum_{i\in [k]} b_i)$.

Moreover, we re-investigate theoretical performance of two existing techniques, \gr and \textsc{Threshold Greedy} (\fast). We proved that: with partition matroid constraint, \gr can obtain a ratio of $\min\big({\alpha} / (1 - (1-\alpha\gamma / \sum_{i \in [k]} b_i)^{\min_{i \in [k]} b_i} ), {(1 + \gamma\alpha)} / \gamma\big)$, which - in comparing with existing work of \cite{friedrich2019greedy} in matroid constraint - has its own advantage in some certain range of non-submodular quantification parameters.

Finally, we investigate our algorithms' performance on two applications of \maxmp: \boostapp and \videoapp. We provide bounds on the objective functions' partition matroid curvature and diminishing ratio to have a better insight on theoretical guarantees of our algorithms. Experimental results show our algorithms return comparable solutions to the state-of-the-art techniques while totally outperform them in the number of queries.

\section{Related Work} \label{sec:related}

\textbf{Quantifying non-submodularity.} To bound how close a function to submodularity, three most popular quantities in literature are: (1) weakly submodular ratio; (2) diminishing return ratio; and (3) generalized curvature. Weakly submodular ratio, denoted as $\gamma_s$, was first introduced by \cite{das2011submodular} and further used by \cite{elenberg2017streaming,qian2015subset,chen2018weakly}. $\gamma_s$ is defined as the maximum value in range $[0,1]$ such that $f(S \cup T) - f(S) \leq \frac{1}{\gamma_s} \sum_{e \in T \setminus S} (f(S \cup \{e\}) - f(S))$ for all $S, T \subseteq V$. Diminishing-return (DR) ratio $\gamma_d$ \cite{bogunovic2018robust,lehmann2006combinatorial,qian2018multiset,kuhnle2018fast} is defined as the largest value in range $[0,1]$ that guarantees $f(T \cup \{e\}) - f(T) \leq \frac{1}{\gamma_d}( f(S \cup \{e\}) - f(S))$ for all $S \subseteq T \subseteq V$ and $e \not\in T$. $\gamma_d$ was proven to be at most the value of $\gamma_s$ \cite{kuhnle2018fast}. General curvature $\alpha_c$ \cite{bian2017guarantees,conforti1984submodular,iyer2013curvature}, on another hand, is the smallest number in $[0,1]$ that $f(T \cup \{e\}) - f(T) \geq (1-\alpha_c)( f(S \cup \{e\}) - f(S))$.

In this work, we adapt DR-ratio and curvature but with more relaxed requirements. To be specific, instead of requiring those quantities applicable for all sets, we narrow down the collection of subsets $S \subseteq T$ that need to satisfy those properties to $|(T \setminus S) \cap V_i| \leq b_i$ for all $i \in [k]$. If considering size constraint, this relaxation is corresponding to the definition of Greedy DR-ratio and Greedy Curvature \cite{bian2017guarantees,kuhnle2018fast}. Not only this relaxation is sufficient to bound our approximation ratios; but also it helps us obtaining meaningful bounds of those quantities in the \maxmp's applications of our experiments.

\textbf{Beyond Cardinality Constraint.} {\em Non-submodular} maximization beyond cardinality constraint has received attention recently. \cite{chen2018weakly} was the first one brought up the concept of non-submodular maximization subject to matroid constraint. In this work, the author proved that \rgr can obtain the ratio of $(1+\frac{1}{\gamma_s})^2$. \cite{gatmiry2018non} then proved \gr is able to obtain a ratio of $\frac{\sqrt{\gamma_s K}  + 1}{0.4 \gamma_s^2}$ and $1+1/\gamma_d$. 

In \textit{submodular} maximization, the study beyond cardinality constraint is too extensive to give a comprehensive overview. Due to space limit, we only go over representative works; and refer readers to comprehensive discussion on \cite{calinescu2011maximizing,buchbinder2019deterministic,friedrich2019greedy}. 

For decades, \gr - with ratio of $2$ \cite{cornnejols1977location} - has been considered as the best algorithm for monotone submodular maximization subject to matroid constraint. This was up until \cite{calinescu2011maximizing} introduced a concept of multilinear extension of submodular functions to devise a $1/(1-1/e)$ algorithm. However, their expensive complexity remains a significant bottleneck to make the algorithm applicable; and how to reduce or improve it is still an intriguing open question for future research. The newest breakthrough is of \cite{buchbinder2019deterministic}, who devised an algorithm, namely \soda, with a ratio of $1/0.5008$ and $\Tilde{O}(nK^2 + KT)$ complexity - where $T$ is the complexity to find a maximum weight perfect matching in a bipartite graph with $2K$ vertices.

The most recent work on partition matroid, to our knowledge, is of \cite{friedrich2019greedy}, in which the authors proved \gr is able to obtain a ratio of ${\alpha_c} / \big(1-\exp{\big[-\alpha_c \frac{\min_{i \in [k]} b_i}{\sum_{i \in [k]} b_i}}\big]\big)$. We generalizes this work to non-submodular objective function by providing analysis that \gr can obtain a ratio of $\min\big({\alpha} / (1 - (1-\alpha\gamma / \sum_{i \in [k]} b_i)^{\min_{i \in [k]} b_i}) , 1/\gamma + \alpha\big)$. If only considering submodular objective function, our ratio has an advantage that it is bounded by $1/\gamma + \alpha$. Therefore, its ratio does not degrade when the input is formed by many partitions.

We also provide approximation ratio of \fast. \fast has been studied by \cite{kuhnle2018fast} in the problem of monotone non-submodular maximization with cardinality constraint. Since partition matroid generalizes cardinality constraint, our analysis techniques are totally different to \cite{kuhnle2018fast}. If projecting our ratio to cardinality constraint, our ratio is better than the one of \cite{kuhnle2018fast}, which is $1 / (1-e^{-\gamma_d \gamma_s (1-\epsilon)} - \epsilon)$. The keys help us obtain a better ratio are (1) $\gamma_s$ is not necessary to bound inequality between obtained solutions and the optimal solution; and (2) we utilizes the general curvature to tighten the inequality equations, thus our ratio becomes better if the curvature moves away from the trivial value $1$.

\section{Definitions and Notations} \label{sec:def}

Given a set function $f$, a set $S$ and $e \not\in S$, denote $\Delta_e f(S) := f(S \cup \{e\}) - f(S)$.

Given the partition matroid constraint of \maxmp, including $V=V_1 \biguplus ... \biguplus V_k$ and $b_1,...,b_k$, denote $b = \sum_{i \in [k]} b_i$; $n = |V|$; $n_i = |V_i|~\forall i \in [k]$. Let $\Bar{n} = \max_{i \in [k]} n_i$ and $\hat{b} = \min_{i \in [k]} b_i$. A set $S \subseteq V$ is called a \textit{maximal} set to the constraint iff $|S \cap V_i| = b_i~\forall i \in [k]$. 


\begin{mydef}
Given an instance of \maxmp, including $V=V_1 \biguplus ... \biguplus V_k; \{b_1,...,b_k\}$ and $f$. The \textbf{Partition Matroid (PM) Diminishing Return} ratio $\gamma$ of the objective function $f$ is defined as the maximum value in $[0,1]$ that guarantees $\Delta_e f(T) \leq \frac{1}{\gamma} \Delta_e f(S)$
for any $S \subseteq T$ that $|(T \setminus S) \cap V_i| \leq b_i ~\forall i \in [k]$ and $e \in V \setminus T$.
\end{mydef}

\begin{mydef}
Given an instance of \maxmp, including $V=V_1 \biguplus ... \biguplus V_k; \{b_1,...,b_k\}$ and $f$. The \textbf{Partition Matroid (PM) Curvature} $\alpha$ of the objective function $f$ is defined as the minimum value in $[0,1]$ that guarantees $ \Delta_e f(T) \geq (1-\alpha) \Delta_e f(S)$ 
for any $S \subseteq T$ that $|(T \setminus S) \cap V_i| \leq b_i ~\forall i \in [k]$ and $e \in V \setminus T$.
\end{mydef}

It is unknown in the literature on how hard it is to obtain exact values of quantities quantifying non-submodularity. $\gamma$ and $\alpha$ are not exception either. Fortunately, for some applications, we can obtain non-trivial bounds of $\gamma$ and $\alpha$, which can help assess approximation ratios of our algorithms. We denote $\gamma^\prime$ as a lower bound of $\gamma$, e.g. $\gamma \geq \gamma^\prime \geq 0$; and $\alpha^\prime$ as a upper bound of $\alpha$, e.g. $\alpha \leq \alpha^\prime \leq 1$.

W.l.o.g, we assume the objective function $f$ is \textit{normalized}, i.e. $f(\emptyset) = 0$, and $b_i \leq n_i$ for all $i \in [k]$. In our algorithms' analysis, we denote $S^*$ as an optimal solution, i.e $f(S^*) = \max_{S :|S \cap V_i| \leq b_i} f(S)$. 





\section{\drand and \fdrand Algorithms} \label{sec:drand}

In this section, we describe \drand, a randomized algorithm with approximation ratio of $(1/\gamma^\prime - 1 +\alpha^\prime) \big(1 - 1/ O(\Bar{n})\big) + 1$. Pseudocode of \drand is presented by Alg. \ref{alg:drand}. In general, \drand works in rounds, and at each round, one member of a group $V_i$ is added to the obtained solution $S$ if $|S \cap V_i| < b_i$. The key for \drand to obtain efficient performance guarantee lies in a random process, which introduces a probability distribution, defined locally for each group, to select a new element of each group to add into the obtained solution (line \ref{line:drand_prob} Alg. \ref{alg:drand}). This random process allows us to construct a sequence of maximal sets in order to form a recursive relationship among changes on the $f$'s values of the obtained solutions, which is critical to bound \drand's approximation ratio.

\begin{algorithm}[t]
    \caption{\drand}
    \label{alg:drand}
    \begin{flushleft}
    \textbf{Input} $V = V_1 \biguplus ... \biguplus V_k; b_1, ..., b_k; f,\gamma^\prime, \alpha^\prime$\\
    \end{flushleft}
    \begin{algorithmic}[1]
        \State $I = [k]; S_0 = \emptyset; t=0$
        \While{$I \neq \emptyset$}
            \For{each $i \in I$} \label{line:inner_loop}
                    \State $a = \ceil{\frac{|V_i \setminus S_t| + 1}{1-\gamma^\prime(1-\alpha^\prime)}} - 1$ 
                    \State $e_t = $ select from $V_i \setminus S_t$ with probability
                    \Statex $\frac{(\Delta_{e_t} f(S_t))^a}{\sum_{u \in V_i \setminus S_t} (\Delta_u f(S_t))^a}$ \label{line:drand_prob}
                    \State $S_{t+1} = S_t \cup \{e_t\}$; $t = t + 1$
                    \If{$|S_t \cap V_i| \geq b_i$} $I = I \setminus \{i\}$
                    \EndIf
            \EndFor
        \EndWhile
    \end{algorithmic}
    \begin{flushleft}
    	\textbf{Return } $S_b$
    \end{flushleft}
\end{algorithm}


\begin{theorem} \label{theorem:drand}
    \drand obtains a $\Big(\frac{1}{\gamma^\prime} + \alpha^\prime - 1 \Big) \Big(1 - \frac{1}{\Bar{n} + 2} \Big) + 1$-approximation solution and has query complexity of $O(\sum_{i \in [k]} n_ib_i)$.
\end{theorem}

\begin{proof}
Denote $\beta = \Big(\frac{1}{\gamma^\prime} + \alpha^\prime - 1 \Big) \Big(1 - \frac{1}{\Bar{n} + 2} \Big)$ and $S_1,...,S_b$ as a sequence of obtained solution by \drand. We prove the approximation ratio of \drand by constructing a sequence of maximal sets $S^*_0,...,S^*_b$ that satisfies the following properties: (1) $S^*_0 = S^*$ and $S^*_b = S_b$; (2) $S_t \subset S^*_t$ for all $t=0,...,b-1$ and $S_b = S^*_b$; (3) $f(S^*_t) - f(S^*_{t+1}) \leq \beta~\mbox{E}\Big[ f(S_{t+1}) - f(S_t) \Big] $ for $t=0 \rightarrow b-1$. Then, we have:
\begin{align*}
    &f(S^*) = \sum_{t=0}^{b-1} \Big( f(S^*_t) - f(S^*_{t+1}) \Big) + f(S^*_b) \\
    &\quad \leq \beta \sum_{t=0}^{b-1} \mbox{E} [ f(S_{t+1}) - f(S_t)] + f(S_b) \leq (\beta + 1)\mbox{E}[f(S_b)]
\end{align*}

To construct the sequence, starting with $S^*_0 = S^*$, for each $t=1,...,b-1$, $S^*_{t+1}$ is formed from $S^*_t$, $S_t$ and $e_t$ as follows: Let $i$ be the index being considered at the \textbf{for} loop (line \ref{line:inner_loop} Alg. \ref{alg:drand}); and $e_t$ will be added into $S_t$. Since $S_t \subset S^*_t$ and $|S_t \cap V_i| < b_i$, $(S^*_t \setminus S_t) \cap V_i \neq \emptyset$. Let $e^\prime$ be any arbitrary element in $(S^*_t \setminus S_t) \cap V_i$.
$S^*_{t+1}$ is set as follows: 
\begin{itemize}[noitemsep,nolistsep]
    \item If $e_t \in (S^*_t \setminus S_t) \cap V_i$, $S^*_{t+1} := S^*_t$. 
    \item Otherwise, let $S^*_{t+1} := S^*_t \setminus \{e^\prime\} \cup \{e_t\}$.  
\end{itemize}
Denote $\rho_e = \Delta_e f(S_t)$ and $\mbox{Pr}_e = \frac{\rho_e^a}{ \sum_{v \in V_i \setminus S_t} \rho_v^a}$ (i.e. $\mbox{Pr}_e$ is probability $e$ is selected). We have:
\begin{align}
    &\mbox{E}\Big[ f(S^*_t) - f(S^*_{t+1}) \Big] \\
    & = \sum_{u \in V_i \setminus S^*_t} \Big[ f(S^*_t) - f(S^*_t \setminus \{e^\prime\} \cup \{u\}) \Big] \times \mbox{Pr}_u \\
    & = \sum_{u \in V_i \setminus S^*_t} \Big[ \Delta_{e^\prime} f(S^*_t \setminus \{e^\prime\}) - \Delta_{u} f(S^*_t \setminus \{e^\prime\}) \Big] \times \mbox{Pr}_u \\
    & \leq \sum_{u \in V_i \setminus S^*_t} \Big[ \frac{1}{\gamma} \rho_{e^\prime} - (1-\alpha) \rho_u \Big] \times \mbox{Pr}_u \label{equ:non_sub_inequal}\\
    & = \frac{1}{\gamma} \sum_{u \in V_i \setminus S^*_t} \frac{\rho_{e^\prime} \rho_u^a}{\sum_{v \in V_i \setminus S_t} \rho_v^a} - (1-\alpha) \sum_{u \in V_i \setminus S^*_t} \rho_u \mbox{Pr}_u \\
    & \leq \frac{1}{\gamma (a + 1)} \sum_{u \in V_i \setminus S^*_t} \frac{\rho_{e^\prime}^{a+1} + a  \rho_u^{a+1}}{\sum_{v \in V_i \setminus S_t} \rho_v^a} \\
    & \quad \quad - (1-\alpha) \sum_{u \in V_i \setminus S^*_t} \rho_u  \mbox{Pr}_u \label{equ:am_gm} \\
    & = \frac{|V_i \setminus S_t^*|}{\gamma(a+1)} \rho_{e^\prime} \mbox{Pr}_{e^\prime} + \Big( \frac{1}{\gamma} \frac{a}{a+1} + \alpha - 1 \Big) \sum_{u \in V_i \setminus S^*_t} \rho_u \mbox{Pr}_u \label{equ:recurse}
\end{align}
where Equ. (\ref{equ:non_sub_inequal}) is from properties of $\gamma$ and $\alpha$; while Equ. (\ref{equ:am_gm}) is from AM-GM inequality.

Replacing $a = \ceil{\frac{|V_i \setminus S_t| + 1}{1 - \gamma^\prime(1-\alpha^\prime)}} - 1$, we have 
\begin{align}
    \frac{|V_i \setminus S_t^*|}{\gamma(a+1)} &\leq \frac{|V_i \setminus S_t|}{\gamma (|V_i \setminus S_t| + 1)/(1 - \gamma^\prime(1-\alpha^\prime))} \\
    & \leq \Big(\frac{1}{\gamma^\prime} + \alpha^\prime - 1 \Big) \Big( 1 - \frac{1}{\Bar{n} + 1} \Big) \label{equ:con_1}
\end{align}
\begin{align}
    \frac{1}{\gamma} \frac{a}{a+1} + \alpha - 1 &\leq \frac{1}{\gamma} \Big( 1 - \frac{1 - \gamma^\prime(1-\alpha^\prime)}{|V_i \setminus S_t| + 2} \Big) + \alpha - 1 \\
    &\leq \Big(\frac{1}{\gamma^\prime} + \alpha^\prime - 1 \Big) \Big( 1 - \frac{1}{\Bar{n} + 2} \Big) \label{equ:con_2}
\end{align}

Therefore, combining Equ. (\ref{equ:con_1}), (\ref{equ:con_2}) to (\ref{equ:recurse}), we have:
\begin{align*}
    (\ref{equ:recurse}) \leq \Big(\frac{1}{\gamma^\prime} + \alpha^\prime - 1 \Big) \Big( 1 - \frac{1}{\Bar{n} + 2} \Big) \mbox{E} \Big[ f(S_{t+1}) - f(S_t) \Big]
\end{align*}
The query complexity of \drand can be trivially inferred from the algorithm's pseudocode.
\end{proof}

\subsection*{How \drand's theoretical performance compares to existing algorithms?} 

Due to differences in definition of the quantities quantifying non-submodularity and how algorithms' ratios depend on them, it is no straight way to compare their ratios. For example, \rgr obtains $(1+\frac{1}{\gamma_s})^2$-ratio \cite{chen2018weakly}. Although $\gamma_s \geq \gamma \geq \gamma^\prime$, it is unclear how this ratio is compared with \drand's ratio. However, \drand has a better query complexity than \rgr ($O(nb)$).  

When $f$ is submodular ($\gamma = 1$), \drand can obtain a ratio of $1 + \alpha^\prime(1 - \frac{1}{\Bar{n} + 2})$. Although \drand's ratio is still not comparable to the best ratio ($1-1/e$) of \cite{calinescu2011maximizing}, their expensive complexity $O(n^8)$ remains a significant bottleneck to make their algorithm applicable in practice. In compare with the most recent work \cite{buchbinder2019deterministic}, \drand can reach a better ratio than \soda ($\frac{1}{0.5008}$) with appropriate values of $\alpha^\prime$ and $\Bar{n}$; and \drand has much better query complexity than \soda ($O(nb^2)$).

\subsection*{Improve \drand's complexity}
\drand's query complexity can be improved by observing that the proof of Theorem \ref{theorem:drand} can non-trivially go through if $e_t$ is selected from a set that overlaps with $(S^*_t \setminus S_t) \cap V_i$ for all $t = 1,...,b$. This always works in Alg. \ref{alg:drand} since $e_t$ is selected from $V_i \setminus S_t$. Therefore, we can use sampling to reduce the space of selecting $e_t$ as in Alg. \ref{alg:fast_drand}.

We call Alg. \ref{alg:fast_drand} \fdrand. The condition, which helps \fdrand has the same ratio as \drand with probability at least $1-\delta$, is guaranteed as stated in the following lemma.
\begin{lemma} \label{lemma:space} $(S^*_t \setminus S_t) \cap R_t \neq \emptyset$ for all $t=0,...,b-1$ with probability at least $1 - \delta$
\end{lemma}
\begin{proof}
We prove for each $t=0,...,b-1$, $\mbox{Pr}\Big[ (S^*_t \setminus S_t) \cap R_t = \emptyset \Big] \leq \frac{\delta}{b}$. Then using union bound, $(S^*_t \setminus S_t) \cap R_t \neq \emptyset$ for all $t=0,...,b -1$ with probability at least $1 - \delta$. This probability is trivial if $R_t = V_i \setminus S_t$. If $|R_t| = \frac{n_i - |S_t \cap V_i|}{b_i - |S_t \cap V_i|} \ln \frac{b}{\delta}$, since $S_t \subseteq S^*_t$, $|(S^*_t \setminus S_t) \cap V_i| = b_i - |S_t \cap V_i|$. We have:
\begin{align*}
    & \mbox{Pr}\Big[ (S^*_t \setminus S_t) \cap R_t = \emptyset \Big] \leq \Big(\frac{|V_i \setminus S^*_t|}{|V_i \setminus S_t|} \Big)^{|R_t|} \\
    & \quad = \Big( 1 - \frac{|(S^*_t \setminus S_t) \cap V_i|}{|V_i \setminus S_t|} \Big)^{|R_t|} \leq e^{- |R_t| \frac{b_i - |S_t \cap V_i|}{n_i - |S_t \cap V_i|}} \leq \frac{\delta}{b}
\end{align*}
which completes the proof.
\end{proof}

\begin{theorem} \label{theorem:drand}
    \fdrand obtains a $	\Big(\frac{1}{\gamma^\prime} + \alpha^\prime - 1 \Big) \Big(1 - \frac{1}{\Bar{n} + 2} \Big) + 1$-approximation solution with probability at least $1-\delta$ and has query complexity of $O(n \ln b \ln \frac{b}{\delta})$.
\end{theorem}

\begin{proof}
Majority proof of \fdrand's approximation ratio overlaps with the proof of \drand. Due to space limit and for the sake of completeness, we provide the proof of \fdrand's ratio in Appendix \ref{apd:drand}. 

In term of query complexity, it is trivial that the number of queries of \fdrand is $\sum_{t=0}^{b-1} |R_t|$. We have:
\begin{align}
    & \sum_{t=0}^{b-1} |R_t| \leq \sum_{i \in [k]} \sum_{j = 0}^{b_i - 1} \frac{n_i-j}{b_i - j} \ln \frac{b}{\delta} \\
    & \quad = \ln \frac{b}{\delta} \Big( \sum_{i \in [k]} b_i + (n_i-b_i) \sum_{j=0}^{b_i - 1} \frac{1}{b_i - j} \Big) \\
    & \quad \leq b \ln \frac{b}{\delta} + \ln \frac{b}{\delta}\sum_{i \in [k]} (n_i-b_i)\ln b_i \leq O(\ln \frac{b}{\delta} \sum_{i \in [k]} n_i \ln b_i) \label{equ:drand_complex} \\
    & \quad \leq O(n \ln \frac{b}{\delta} \ln \sum_{i \in [k]} \frac{n_ib_i}{n} ) \leq O(n \ln b \ln \frac{b}{\delta}) \label{equ:drand_concave}
\end{align}
where Equ. (\ref{equ:drand_concave}) is from the fact that $\log x$ is a concave function, so $\sum_{i} \alpha_i \log x_i \leq \log \alpha_i x_i$ if $\sum_i \alpha_i = 1$; and $\sum_{i \in [k]} \frac{n_i b_i}{n} \leq \sum_{i \in [k]} \frac{n_i}{n} \sum_{i \in [k]} b_i = b$.
\end{proof}

\begin{algorithm}[t]
            \caption{\fdrand}
            \label{alg:fast_drand}
            \begin{flushleft}
            \textbf{Input} $V = V_1 \biguplus ... \biguplus V_k; f, \gamma^\prime, \alpha^\prime; b_1, ..., b_k; \delta \in [0,1]$\\
            \end{flushleft}
            \begin{algorithmic}[1]
                \State $I = [k]; S_0 = \emptyset; t=0$
                \While{$I \neq \emptyset$}
                    \For{each $i \in I$} \label{line:inner_loop}
                        \State $R_t = $ pick $\min\big(\frac{n_i - |S_t \cap V_i|}{b_i - |S_t \cap V_i|} \ln \frac{b}{\delta}, |V_i \setminus S_t| \big)$ 
                        \Statex random elements from $V_i \setminus S_t$
                        \State $a = \ceil{\frac{|R_t| + 1}{1-\gamma^\prime(1-\alpha^\prime)}} - 1$ 
                        \State $e_t = $ select from $R_t$ with probability
                        \Statex $\frac{(\Delta_{e_t} f(S_t))^a}{\sum_{u \in R_t} (\Delta_u f(S_t))^a}$ \label{line:drand_prob}
                        \State $S_{t+1} = S_t \cup \{e_t\}$; $t = t + 1$
                        \If{$|S_t \cap V_i| \geq b_i$} $I = I \setminus \{i\}$
                        \EndIf
                    \EndFor
                \EndWhile
            \end{algorithmic}
            \begin{flushleft}
            	\textbf{Return } $S_b$
            \end{flushleft}
        \end{algorithm}
\section{\gr-like Algorithms} \label{sec:fast}


We re-study the theoretical performance guarantee of two algorithms, \gr and \fast. Our analysis provides better ratios of \gr than existing works on matroid constraint \cite{gatmiry2018non} or submodular objective function \cite{friedrich2019greedy}. 

In general, \gr works in round and at each round, an element of maximal marginal gain, whose addition does not violate partition matroid constraint, is added to the obtained solution. The algorithm terminates when the obtained solution is maximal. \fast, on the other hand, works by always keeping a threshold $\tau$, which bounds the maximum marginal gain to the objective by any non-selected elements. The algorithm runs in rounds; at each round, any element with a marginal gain at least $\tau$ will be added to the solution if it does not violate the partition matroid constraint. After each round, $\tau$ is decreased by a factor $1-\epsilon$ in order to guarantee new elements can be added to the solution at successive rounds. The algorithm continues until the obtained solution becoming a maximal set or the threshold is below a value defined by $\epsilon$ and $b$. \gr's pseudocode is presented by Alg. \ref{alg:greedy} and \fast's is Alg. \ref{alg:fast}.

\begin{theorem} \label{theorem:greedy}
    \gr obtains a $\min(\frac{1}{r^{(g)}_1}, \frac{1}{r^{(g)}_2})$-approximation solution, where
    \begin{align*}
        r^{(g)}_1 = \frac{\gamma}{1 + \gamma\alpha} \quad\quad
        r^{(g)}_2 = \frac{1}{\alpha} \Big[ 1 - \Big( 1 - \frac{\alpha\gamma}{b} \Big)^{\hat{b}} \Big]
    \end{align*}
    and has a query complexity of $O(n~b)$.
\end{theorem}
      
\begin{theorem} \label{theorem:fast}
    \fast obtains a $\min(\frac{1}{r^{(t)}_1}, \frac{1}{r^{(t)}_2})$-approximation solution, where
    \begin{align*}
        r^{(t)}_1 = \frac{\gamma(1-\epsilon)^2}{1 + \gamma\alpha(1-\epsilon)} \quad
        r^{(t)}_2 = \frac{1}{\alpha} \Big[ 1 - \Big( 1 - \frac{\alpha\gamma(1-\epsilon)}{b} \Big)^{\hat{b}} \Big]
    \end{align*}
    and has a query complexity of $O(\frac{n}{\epsilon} \ln b)$.
\end{theorem}

\begin{algorithm}[t]
            \caption{\gr}
            \label{alg:greedy}
            \begin{flushleft}
            \textbf{Input} $V = V_1 \biguplus ... \biguplus V_k; f; b_1, ..., b_k$\\
            \end{flushleft}
            \begin{algorithmic}[1] 
                \State $I = [k]; S_0 = \emptyset; t=0$
                \While{$I \neq \emptyset$}
                    \State $e, i = \mbox{argmax}_{e \in V_i \setminus S_t; i \in I} \Delta_e f(S_t)$
                    \State $S_{t+1} = S_t \cup \{e\}$; $t = t+1$
                    \If{$|S_t \cap V_i| \geq b_i$}  $I = I \setminus \{i\}$
                    \EndIf
                \EndWhile
            \end{algorithmic}
            \begin{flushleft}
            	\textbf{Return } $S_b$
            \end{flushleft}
        \end{algorithm}

Due to space limit, the full proof of Theorem \ref{theorem:greedy} and \ref{theorem:fast} is provided in Appendix \ref{apd:fast}.

In case of submodular objective function, $r_2^{(g)}$ of \gr is identical to the ratio obtained by \cite{friedrich2019greedy}. With cardinality constraint, $r_2^{(g)}$ matches with the ratio of \cite{bian2017guarantees}, which was also proven to be tight. However, with $\hat{b}/b \rightarrow 0$ (e.g. the input is formed by many partitions), $r_2^{(g)}$ and $r_2^{(t)}$ approach 0 and become undesirable. In this case, $r_1^{(g)}$ and $r_1^{(t)}$ should be a better bound on the performance of \gr and \fast. 

\begin{algorithm}[t]
            \caption{\fast}
            \label{alg:fast}
            \begin{flushleft}
            \textbf{Input} $V = V_1 \biguplus ... \biguplus V_k; f; b_1, ..., b_k; \epsilon \in [0,1]$\\
            \end{flushleft}
            \begin{algorithmic}[1] 
                \State $I = [k]; S_0 = \emptyset; t=0$
                \State $\tau = \tau_0 = \max_{e \in V} \Delta_e f(S_0)$ 
                \While{$I \neq \emptyset$ and $\tau \geq \frac{\epsilon (1-\epsilon) \tau_0}{b}$} \label{line:fast_loop}
                    \For{each $i \in I$ and $e \in V_i \setminus S_t$}
                            \If{$\Delta_{e} f(S_t) \geq \tau$} \label{line:fast_con}
                                \State $S_{t+1} = S_t \cup \{e\}$; $t = t + 1$
                                \If{$|S_t \cap V_i| \geq b_i$} $I = I \setminus \{i\}$
                                \EndIf
                            \EndIf                   
                    \EndFor
                    \State $\tau = \tau (1-\epsilon)$
                \EndWhile
            \end{algorithmic}
            \begin{flushleft}
            	\textbf{Return } $S_t$
            \end{flushleft}
        \end{algorithm}

\section{Applications and Experimental Results} \label{sec:experiment}

In this section, we consider two applications of \maxmp: \boostapp and \videoapp.

\textbf{\boostapp}. In this problem, a social directed graph $G=(V,E)$ is given, where $V$ represents a set of social network users; and $E$ represents friendship between social users in $V$. An information will start spreading at a set $I \subset V$ of users. The problem asks for a set $S$ of users to strengthen the influence spread in order to maximize the number of users the information can reach.

\boostapp under size constraint has been studied by \cite{lin2017boosting}. In their model, each edge $e=(u,v) \in E$ is associated with two weight values $p_e^0, p_e^1$ ($p_e^0 \leq p_e^1 \leq 1$). The probability $v$ adopts the information from $u$ is $p_e^1$ if $v \in S$; $p_e^0$ otherwise. In this application, $f(S)$ measures expected number of users the information can reach if $S$ is selected. The authors has proven that $f$ is monotone non-submodular; but did not show how close $f$ is to submodularity. We provide the bound $\gamma^\prime, \alpha^\prime$ of $\gamma$ and $\alpha$ of $f$ as in Lemma \ref{lemma:boost_bound}, and full proof is provided in Appendix \ref{apd:exp}.
\begin{lemma} \label{lemma:boost_bound}
    Given a \boostapp instance, let $\Delta$ be the maximum in-degree of the input directed graph. For any $S \subseteq T$ that $|(T \setminus S) \cap V_i| \leq b_i ~\forall i \in [k]$ and $u \in V \setminus T$:
    \begin{align}
        \min_{|E^\prime| \leq b\Delta} \prod_{e \in E^\prime} \frac{1 - p_{e}^1}{1 - p_{e}^0} &\leq \frac{\Delta_u f(S)}{\Delta_u f(T)} \leq \max_{|E^\prime| \leq b\Delta} \prod_{e \in E^\prime} \frac{p_e^1}{p_e^0}
    \end{align}
\end{lemma}

\textbf{\videoapp} Given a video, this application aims to pick a few representative frames from the video which can contains as much content as possible. The video contains $n$ frames; each frame is represented by a $p$-dimensional vector. Let $X \in \mathbb{R}^{n\times n}$ be the Gramian matrix of the $n$ resulting vectors and the Gaussian kernel; i.e. $X_{ij}$ is the value of the Gaussian kernel between the $i$-th and $j$-th vectors. The objective function is defined as $f(S) = \mbox{det}(I + X_S)$, where $X_S$ is the submatrix of $X$ indexed by $S$; and $I$ is a unit matrix. 

$f(S)$ was proved to be supermodular by \cite{bian2017guarantees}, thus its curvature $\alpha = 0$. The authors also bounded the weakly submodular ratio, which is not useful in our algorithms. We  bound the value of $\gamma$ as in the following lemma, and full proof is provided in Appendix \ref{apd:exp}.

\begin{lemma} \label{lemma:video_bound}
Given a \videoapp instance, let $A = I + X$ and $\lambda_i(M)$ be the $i$-th eigenvalue of a positive definite matrix $M$ in a way that $\lambda_1(M) \geq ... \geq \lambda_{\mbox{rank}(M)}(M)$. For any $S \subseteq T$ that $|(T \setminus S) \cap V_i| \leq b_i ~\forall i \in [k]$ and $e \in V \setminus T$:
    \begin{align}
        \Delta_e f(S) \geq \Delta_e f(T) \times \frac{ \lambda_n(A) - 1 }{  \lambda_1(A) - 1 } \prod_i^{b} \frac{1}{\lambda_i(A)} 
    \end{align}
\end{lemma}

\begin{figure}[H]
\centering
\begin{tikzpicture}[yscale=0.6, xscale=0.7]
    \begin{groupplot}[group style={group size= 2 by 2}, width=.38\textwidth]
        \nextgroupplot[xlabel={$b$}, title={Influence Spread}, grid style=dashed, grid=both, grid style={line width=.1pt, draw=gray!10}, major grid style={line width=.2pt,draw=gray!50}, every axis plot/.append style={ultra thick, smooth}, every x tick label/.append style={font=\scriptsize}, every y tick label/.append style={font=\scriptsize}, xtick distance={20}, ytick distance={50},]
                \addplot[black,mark=o] table [x=B, y=Soda, col sep=comma] {data/facebook/budget/solution.csv}; \label{plot:soda}
                \addplot[green,mark=o] table [x=B, y=Greedy, col sep=comma] {data/facebook/budget/solution.csv}; \label{plot:greedy}
                \addplot[orange,mark=o] table [x=B, y=Random_Greedy, col sep=comma] {data/facebook/budget/solution.csv}; \label{plot:random_greedy}
                \addplot[blue,mark=o] table [x=B, y=Randomize, col sep=comma] {data/facebook/budget/solution.csv}; \label{plot:randomize}
                \addplot[red,mark=square] table [x=B, y=Fast_5, col sep=comma] {data/facebook/budget/solution.csv}; \label{plot:fast_2}
                \coordinate (top) at (rel axis cs:0,1);
        \nextgroupplot[xlabel={$b$}, title={\# queries}, grid style=dashed, grid=both, grid style={line width=.1pt, draw=gray!10}, major grid style={line width=.2pt,draw=gray!50}, every axis plot/.append style={ultra thick, smooth}, every x tick label/.append style={font=\scriptsize}, every y tick label/.append style={font=\scriptsize}, xtick distance={20}]
                \addplot table [x=B, y=Randomize, col sep=comma] {data/facebook/budget/query.csv};
                \addplot table [x=B, y=Fast_5, col sep=comma] {data/facebook/budget/query.csv};
                \addplot[green,mark=o] table [x=B, y=Greedy, col sep=comma] {data/facebook/budget/query.csv};
                \addplot[orange,mark=o] table [x=B, y=Random_Greedy, col sep=comma] {data/facebook/budget/query.csv};
                \addplot[black,mark=o] table [x=B, y=Soda, col sep=comma] {data/facebook/budget/query.csv};
        \nextgroupplot[xlabel={$k$}, title={Influence Spread}, grid style=dashed, grid=both, grid style={line width=.1pt, draw=gray!10}, major grid style={line width=.2pt,draw=gray!50}, every axis plot/.append style={ultra thick, smooth}, every x tick label/.append style={font=\scriptsize}, every y tick label/.append style={font=\scriptsize}, xtick distance={4}, yshift=-0.5cm]
                \addplot table [x=num_groups, y=Randomize, col sep=comma] {data/facebook/num_groups/solution.csv};
                \addplot table [x=num_groups, y=Fast_5, col sep=comma] {data/facebook/num_groups/solution.csv};
                \addplot[green,mark=o] table [x=num_groups, y=Greedy, col sep=comma] {data/facebook/num_groups/solution.csv};
                \addplot[orange,mark=o] table [x=num_groups, y=Random_Greedy, col sep=comma] {data/facebook/num_groups/solution.csv}; 
                \addplot[black,mark=o] table [x=num_groups, y=Soda, col sep=comma] {data/facebook/num_groups/solution.csv};
         \nextgroupplot[xlabel={$k$}, title={\# queries}, grid style=dashed, grid=both, grid style={line width=.1pt, draw=gray!10}, major grid style={line width=.2pt,draw=gray!50}, every axis plot/.append style={ultra thick, smooth}, every x tick label/.append style={font=\scriptsize}, every y tick label/.append style={font=\scriptsize}, xtick distance={4}, ymode=log, yshift=-0.5cm]
                \addplot table [x=num_groups, y=Randomize, col sep=comma] {data/facebook/num_groups/query.csv};
                \addplot table [x=num_groups, y=Fast_5, col sep=comma] {data/facebook/num_groups/query.csv};
                \addplot[green,mark=o] table [x=num_groups, y=Greedy, col sep=comma] {data/facebook/num_groups/query.csv};
                \addplot[orange,mark=o] table [x=num_groups, y=Random_Greedy, col sep=comma] {data/facebook/num_groups/query.csv};
                \addplot[black,mark=o] table [x=num_groups, y=Soda, col sep=comma] {data/facebook/num_groups/query.csv};
                \coordinate (bot) at (rel axis cs:1,0);
    \end{groupplot}
\matrix[
    matrix of nodes,
    anchor=south,
    font=\tiny,
    xshift=4cm,
    yshift=3cm
  ]
  {
    \ref{plot:randomize} \fdrand ~
    \ref{plot:fast_2} \fast ~
    \ref{plot:greedy} \gr \\
    \ref{plot:random_greedy} \rgr ~
    \ref{plot:soda} \soda\\};
\end{tikzpicture}
\caption{Performance in \boostapp.}
 	\label{fig:boost}
\end{figure}

\begin{figure}[H]
\centering
\begin{tikzpicture}[yscale=0.6, xscale=0.7]
    \begin{groupplot}[group style={group size= 2 by 2}, width=.38\textwidth]
        \nextgroupplot[xlabel={$b$}, title={Information Gain}, grid style=dashed, grid=both, grid style={line width=.1pt, draw=gray!10}, major grid style={line width=.2pt,draw=gray!50}, every axis plot/.append style={ultra thick, smooth}, every x tick label/.append style={font=\scriptsize}, every y tick label/.append style={font=\scriptsize}]
                \addplot[black,mark=o] table [x=B, y=Soda, col sep=comma] {data/cooking/budget/solution.csv};
                \addplot[green,mark=o] table [x=B, y=Greedy, col sep=comma] {data/cooking/budget/solution.csv};
                \addplot[orange,mark=o] table [x=B, y=Random_Greedy, col sep=comma] {data/cooking/budget/solution.csv};
                \addplot[blue,mark=o] table [x=B, y=Randomize, col sep=comma] {data/cooking/budget/solution.csv};
                \addplot[red,mark=square] table [x=B, y=Fast_5, col sep=comma] {data/cooking/budget/solution.csv};
                \coordinate (top) at (rel axis cs:0,1);
        \nextgroupplot[xlabel={$b$}, title={\# queries}, grid style=dashed, grid=both, grid style={line width=.1pt, draw=gray!10}, major grid style={line width=.2pt,draw=gray!50}, every axis plot/.append style={ultra thick, smooth}, every x tick label/.append style={font=\scriptsize}, every y tick label/.append style={font=\scriptsize}]
                \addplot table [x=B, y=Randomize, col sep=comma] {data/cooking/budget/query.csv};
                \addplot table [x=B, y=Fast_5, col sep=comma] {data/cooking/budget/query.csv};
                \addplot[green,mark=o] table [x=B, y=Greedy, col sep=comma] {data/cooking/budget/query.csv};
                \addplot[orange,mark=o] table [x=B, y=Random_Greedy, col sep=comma] {data/cooking/budget/query.csv};
                \addplot[black,mark=o] table [x=B, y=Soda, col sep=comma] {data/cooking/budget/query.csv};
        \nextgroupplot[xlabel={$k$}, title={Information Gain}, grid style=dashed, grid=both, grid style={line width=.1pt, draw=gray!10}, major grid style={line width=.2pt,draw=gray!50}, every axis plot/.append style={ultra thick, smooth}, every x tick label/.append style={font=\scriptsize}, every y tick label/.append style={font=\scriptsize}, xtick distance={4}, yshift=-0.5cm]
                \addplot table [x=num_groups, y=Randomize, col sep=comma] {data/cooking/num_groups/solution.csv};
                \addplot table [x=num_groups, y=Fast_5, col sep=comma] {data/cooking/num_groups/solution.csv};
                \addplot[green,mark=o] table [x=num_groups, y=Greedy, col sep=comma] {data/cooking/num_groups/solution.csv};
                \addplot[orange,mark=o] table [x=num_groups, y=Random_Greedy, col sep=comma] {data/cooking/num_groups/solution.csv}; 
                \addplot[black,mark=o] table [x=num_groups, y=Soda, col sep=comma] {data/cooking/num_groups/solution.csv};
         \nextgroupplot[xlabel={$k$}, title={\# queries}, grid style=dashed, grid=both, grid style={line width=.1pt, draw=gray!10}, major grid style={line width=.2pt,draw=gray!50}, every axis plot/.append style={ultra thick, smooth}, every x tick label/.append style={font=\scriptsize}, every y tick label/.append style={font=\scriptsize}, xtick distance={4}, ymode=log, yshift=-0.5cm]
                \addplot table [x=num_groups, y=Randomize, col sep=comma] {data/cooking/num_groups/query.csv};
                \addplot table [x=num_groups, y=Fast_5, col sep=comma] {data/cooking/num_groups/query.csv};
                \addplot[green,mark=o] table [x=num_groups, y=Greedy, col sep=comma] {data/cooking/num_groups/query.csv};
                \addplot[orange,mark=o] table [x=num_groups, y=Random_Greedy, col sep=comma] {data/cooking/num_groups/query.csv};
                \addplot[black,mark=o] table [x=num_groups, y=Soda, col sep=comma] {data/cooking/num_groups/query.csv};
                \coordinate (bot) at (rel axis cs:1,0);
    \end{groupplot}
\end{tikzpicture}
\caption{Performance in \videoapp}
 	\label{fig:video}
\end{figure}

\subsection{Settings and Compared Algorithms}

With \boostapp, we use Facebook dataset from SNAP database \cite{snapnets}, an undirected graph with 4,039 nodes and 88,234 edges. Since it is undirected, we treat each edge as two directed edges. For each edge $e=(u,v)$, $p_e^0 = \frac{1}{d_v}$ and $p_e^1 = \frac{2}{d_v}$ where $d_v$ is in-degree of $v$. Information starts spreading at a node of highest degree. Due to lack of information, a user is randomly assigned to a group $V_i$. The budget is distributed equally to each group, i.e. $b_1 \approx ... \approx b_k \approx \frac{b}{k}$. The objective is estimated over 100 pre-sampled graph realizations of $G$.

With \videoapp, we chose a video of roughly 3.5 minutes. The video is segmented to $k$ equal-length parts; and the algorithms will pick $\frac{b}{k}$ frames from each part.

With \fdrand, we set $\delta=0.001$, which guarantees \fdrand to return solutions almost similar to \drand but be much better in the number of queries. With \fast, we set $\epsilon = 0.5$.  Results were averaged over 10 repetitions.

We varied values of $b$ and $k$; and compare \fdrand, \gr and \fast with \rgr \cite{chen2018weakly} and \soda \cite{buchbinder2019deterministic}. Although \soda's performance is unknown if $f$ is submodular, we used it as a heuristic to compare. Source code is available at \url{https://github.com/lannn2410/maxmp}.

\subsection{Numerical Results} 

Fig. \ref{fig:boost} and \ref{fig:video} show experimental results of different algorithms on \boostapp and \videoapp. With experiments that we varied values of $b$, we fixed $k=2$. With the one that $k$ is varied, we fixed $b=100$ in \boostapp and $b=20$ in \videoapp.

In these experiments, \fdrand, \gr and \soda performed approximately equal in term of solution quality while \fast was always the worst one. Especially, in \videoapp, the supermodular objective function made the marginal gain of non-included elements increase with larger obtained solutions. Therefore, \fast easily reached a maximal solution just by one or two iterations of decreasing threshold. That explained why \fast took very few number of queries but has undesirable returned solution quality. In term of the number of queries, \fdrand outperformed \gr, \rgr and \soda. 




\fdrand closed the gap or even surpassed \fast to become the best algorithm in the number of queries in the experiments with fixed $b$ and varied $k$. In these experiments, we can see that the number of queries of all algorithms, except \fdrand, almost did not change or just slightly decreased with larger $k$. \fdrand's numbers, on the other hand, decreased significantly as $k$ increased. This phenomenon is also reflected on the theoretical bound of \fdrand's complexity. In Equ. (\ref{equ:drand_complex}), \fdrand's complexity is bounded by $ O(\ln \frac{b}{\delta} \sum_{i \in [k]} n_i \ln b_i)$. With $n_i$s are roughly equal (the same with $b_i$s), \fdrand's complexity becomes $O(n\ln \frac{b}{k} \ln \frac{b}{\delta})$, which decreases w.r.t $k$.

\section{Discussion} \label{sec:conclusion}
We proposed \drand and later \fdrand to solve monotone non-submodular maximization with partition matroid constraint. The experimental results demonstrated that \fdrand can perform closely to the best algorithms in solution quality, and  outperform other algorithms (except \fast - the worst in solution quality) in the number of queries. Although there is no superior algorithm in general, \fdrand should be considered as the best algorithm in scenarios that scalability issues are concerned, e.g. algorithms with fast runtime and relatively high solution quality.

There is still an open question on what is the best algorithm in approximation ratio? \drand's ratio depends on $\gamma^\prime, \alpha^\prime$ - which can be undesirable in some settings of our experiments. However, it is unknown on how hard to obtain exact value of $\gamma, \alpha$ or other non-submodular quantities. And it is too expensive for us if computing those quantities by enumerating all possible $S,T$ that $T \setminus S$ satisfies partition matroid. Therefore, it is still open on how different between \gr, \fast, \rgr and \drand's ratio.

\section*{Acknowledgements}\label{sc:acknowledgements}
This work was supported in part by the National Science Foundation (NSF) grants IIS-1908594, CNS-1814614. We would like to thank the anonymous reviewers for their helpful feedback.

\bibliographystyle{named}
\bibliography{reference}
\clearpage
\input{appendix}



\end{document}

%% file: appendix.tex
\onecolumn
\appendix





\section{Proof of \fdrand's Ratio} \label{apd:drand}

Similar to \drand, we prove that: In \fdrand, with high probability (i.e. $1-\delta$) there exists a way to construct a sequence of maximal sets $S^*_0,...,S^*_b$ that satisfies (1) $S^*_0 = S^*$ and $S^*_b = S_b$; (2) $S_t \subseteq S^*_t$ for all $t=0,...,b$; and (3) $f(S^*_t) - f(S^*_{t+1}) \leq \beta~\mbox{E}\Big[ f(S_{t+1}) - f(S_t) \Big] $. If these three properties are guaranteed, the ratio of \fdrand follows.

For each $t=0,...,b-1$, a set $S^*_{t+1}$ is formed from $S^*_t$, $S_t$ and $e_t$ but with a condition that $(S^*_t \setminus S_t) \cap R_t \neq \emptyset$, which is guaranteed with high probability by Lemma \ref{lemma:space}. 

Let's consider a moment when the $t$-th element is added into the solution. Let $i$ be the index being considered when $e_t$ is added into $S_{t-1}$. With $(S^*_t \setminus S_t) \cap R_t \neq \emptyset$, let $e^\prime$ be an arbitrary element in $(S^*_t \setminus S_t) \cap R_t$, then $S^*_{t+1}$ is set as follows: 
\begin{itemize}
    \item If $e_t \in (S^*_t \setminus S_t) \cap R_t$, $S^*_{t+1} := S^*_t$
    \item Otherwise, $S^*_{t+1} := S^*_t \setminus \{e^\prime\} \cup \{e_t\}$
\end{itemize}

Then we have:
\begin{align}
    \mbox{E}\Big[ f(S^*_t) - f(S^*_{t+1}) \Big] &= \sum_{u \in R_t \setminus S^*_t} \Big[ f(S^*_t) - f(S^*_t \setminus \{e^\prime\} \cup \{u\}) \Big] \times \mbox{Pr}_u \\
    & = \sum_{u \in R_t \setminus S^*_t} \Big[ \Delta_{e^\prime} f(S^*_t \setminus \{e^\prime\}) - \Delta_{u} f(S^*_t \setminus \{e^\prime\}) \Big] \times \mbox{Pr}_u \\
    & \leq \sum_{u \in R_t \setminus S^*_t} \Big[ \frac{1}{\gamma} \rho_{e^\prime} - (1-\alpha) \rho_u \Big] \times \mbox{Pr}_u \label{equ:fnon_sub_inequal}\\
    & = \frac{1}{\gamma} \sum_{u \in R_t \setminus S^*_t} \frac{\rho_{e^\prime} \rho_u^a}{\sum_{v \in R} \rho_v^a} - (1-\alpha) \sum_{u \in R_t \setminus S^*_t} \rho_u \mbox{Pr}_u \\
    & \leq \frac{1}{\gamma (a + 1)} \sum_{u \in R_t \setminus S^*_t} \frac{\rho_{e^\prime}^{a+1} + a \times \rho_u^{a+1}}{\sum_{v \in R} \rho_v^a} - (1-\alpha) \sum_{u \in R_t \setminus S^*_t} \rho_u \times \mbox{Pr}_u \label{equ:fam_gm} \\
    & = \frac{|R_t \setminus S_t^*|}{\gamma(a+1)} \rho_{e^\prime} \mbox{Pr}_{e^\prime} + \Big( \frac{1}{\gamma} \frac{a}{a+1} + \alpha - 1 \Big) \sum_{u \in R_t \setminus S^*_t} \rho_u \mbox{Pr}_u \label{equ:frecurse}
\end{align}
where Equ. (\ref{equ:fnon_sub_inequal}) is from properties of $\gamma$ and $\alpha$; while Equ. (\ref{equ:fam_gm}) is from AM-GM inequality.


With $a = \ceil{\frac{|R_t| + 1}{1 - \gamma^\prime(1-\alpha^\prime)}} - 1$, we have 
\begin{align*}
    &\frac{|R_t \setminus S_t^*|}{\gamma(a+1)} \leq \frac{|R_t|}{\gamma \frac{|R_t| + 1}{1 - \gamma^\prime(1-\alpha^\prime)}} \leq \Big(\frac{1}{\gamma^\prime} + \alpha^\prime - 1 \Big) \Big( 1 - \frac{1}{\Bar{n} + 1} \Big) \\
    &\frac{1}{\gamma} \frac{a}{a+1} + \alpha - 1 \leq \frac{1}{\gamma} \Big( 1 - \frac{1 - \gamma^\prime(1-\alpha^\prime)}{|R_t| + 2} \Big) + \alpha - 1 \leq \Big(\frac{1}{\gamma^\prime} + \alpha^\prime - 1 \Big) \Big( 1 - \frac{1}{\Bar{n} + 2} \Big)
\end{align*}

Therefore, 
\begin{align*}
    (\ref{equ:frecurse}) \leq \Big(\frac{1}{\gamma^\prime} + \alpha^\prime - 1 \Big) \Big( 1 - \frac{1}{\Bar{n} + 2} \Big) \mbox{E} \Big[ f(S_{t+1}) - f(S_t) \Big]
\end{align*}
which completes the proof.

\section{Proofs of \gr \& \fast} \label{apd:fast}

We use a common framework that can be used to prove approximation ratio of both \gr and \fast. Denote $e_0, e_1, ..., e_{t-1}$ as a sequence of elements added to the obtained solution, i.e. $S_j = \{e_0,e_1, ..., e_{j-1}\}$. In \fast, we assume $\tau_0 =\max_{e \in V} \Delta_e f(S_0) \leq \gamma f(S^*)$ since if not, the ratio can be obtained trivially by: $f(S_t) \geq \Delta_{e_0} f(S_0) \geq \gamma f(S^*)$.

The reason we use a common proof for \gr and \fast because both algorithms guarantee: For each $j=0,...,t-1$, $\Delta_{e_j} f(S_{j}) \geq \gamma\sigma \Delta_u f(T)$ for any $T \supseteq S_{j}$ and $u \in V_i \setminus S_{j}$ that $|S_{j} \cap V_i| < b_i$, where $\sigma = 1$ in \gr and $1-\epsilon$ in \fast. This guarantee is trivial with \gr due to PM DR-ratio $\gamma$'s property. To prove this guarantee in \fast, we observe that: since $u \not\in S_{j}$, there should exist $j^\prime \leq j$ that $\Delta_u f(S_{j^\prime}) \leq \frac{\tau_j}{1-\epsilon} \leq \frac{\Delta_{e_j} f(S_j)}{1-\epsilon}$, where $\tau_j$ is the threshold $\tau$ when $e_j$ is added. And because $S_{j^\prime} \subseteq S_{j} \subseteq T$, $\Delta_u f(S_{j^\prime}) \geq \gamma \Delta_u f(T)$.

For simplicity, in the common proof, we denote $r_1$ as $r^{(g)}_1$ (for \gr) and $r^{(t)}_1$ (for \fast). The same notation is applied for $r_2$. The ratios of two algorithms means that: (1) $f(S_t) \geq r_1 f(S^*)$ and (2) $f(S_t) \geq r_2 f(S^*)$. Therefore, our proof focuses on proving these two statements.
\begin{lemma}
$f(S_t) \geq r_1 f(S^*)$
\end{lemma}
\begin{proof}
The lemma is proven by constructing a sequence of maximal sets $S^*_0,...,S^*_t$ as follows: Starting with $S^*_0 = S^*$. Assuming $i \in I$ is the index being consider when the algorithm adds $e_{j}$ into the current obtained solution $S_j$, pick an arbitrary $u \in S^*_j \setminus S_j \cap V_i$. if $u = e_{j}$ then $S^*_{j+1} := S^*_j$ and we have $f(S^*_j) - f(S^*_{j+1}) = 0$. Otherwise,  set $S^*_{j+1} := S^*_j \setminus \{u\} \cup \{e_{j}\}$, we have: 
\begin{align*}
        f(S^*_j) - f(S^*_{j+1}) &= \Delta_u f(S^*_j \setminus \{u\}) - \Delta_{e_{j}} f(S^*_j \setminus \{u\})  \leq \Big( \frac{1}{\gamma\sigma} + \alpha - 1 \Big) \Delta_{e_{j}} f(S_j)
\end{align*}
Therefore, from the sequence $S^*_0, ..., S^*_t$, we have:
\begin{align}
    &f(S^*) - f(S^*_t) = \sum_{j=0}^{t-1} \Big( f(S^*_j) - f(S^*_{j+1}) \Big)  \leq \Big( \frac{1}{\gamma\sigma} + \alpha - 1 \Big) \sum_{j=0}^{t-1} \Big[ f(S_{j+1}) - f(S_j) \Big]  \leq \Big( \frac{1}{\gamma\sigma} + \alpha - 1 \Big) f(S_t) \label{equ:fast_recurse}
\end{align}
With \gr, the lemma follows from Equ. (\ref{equ:fast_recurse}) since in \gr, $t=b$ and $S^*_b = S_b$. This is also applied to \fast if the returned $S_t$ is maximal. Otherwise, denote $\{e_1^\prime, ..., e_{b-t}^\prime\}$ as a sequence of elements in $S^*_t \setminus S_t$. Then for each $e_l^\prime \in S^*_t \setminus S_t$, there exists $j \leq t$ such that $S_j \subseteq S_t \subseteq S_t \cup \{e^\prime_1,...,e^\prime_{l-1}\}$ and $\Delta_{e^\prime_l} f(S_t \cup \{e^\prime_1,...,e^\prime_{l-1}\}) \leq \frac{1}{\gamma}\Delta_{e^\prime_l} f(S_j) \leq \frac{\epsilon \tau_0}{\gamma b}$. Therefore,

\begin{align}
    f(S^*_t) - f(S_t) &= \sum_{l=1}^{b-t} \Delta_{e^\prime_l} f(S_t \cup \{e^\prime_1,...,e^\prime_{l-1}\})  \leq b \frac{\epsilon \tau_0}{\gamma b} \leq \epsilon f(S^*) \label{equ:fast_less}
\end{align}
Combining Equ. (\ref{equ:fast_recurse}) and (\ref{equ:fast_less}), we complete the lemma in the \fast case.
\end{proof}
\begin{lemma} \label{lemma:r_2}
$f(S_t) \geq r_2 f(S^*)$
\end{lemma}


\begin{proof}

Denote $S^* = \{e^*_1,...,e^*_b\}$. We consider two cases that can happen when the algorithms terminates: (1) $|S_t \cap V_i| < b_i$ for all $i \in [k]$; and (2) There exists $i \in [k]$ that $|S_t \cap V_i| = b_i$.

Case (1) never happens with \gr. If it happens with \fast, then for each $e^*_i \in S^* \setminus S$, there should exists $j \leq t$ such that $S_j \subseteq S \subseteq S \cup \{e^*_1,...,e^*_{i-1}\}$ and $\Delta_{e^*_i} f(S \cup \{e^*_1,...,e^*_{i-1}\}) \leq \frac{1}{\gamma}\Delta_{e^*_i} f(S_j) \leq \frac{\epsilon \tau_0}{\gamma b}$. We have:
\begin{align*}
    f(S^*) - f(S) & \leq f(S^* \cup S) - f(S) = \sum_{e^*_i \in S^*} \Delta_{e^*_i} f(S \cup \{e^*_1,...,e^*_{i-1}\}) \\
    & = \sum_{e^*_i \in S^* \setminus S} \Delta_{e^*_i} f(S \cup \{e^*_1,...,e^*_{i-1}\}) \leq b \frac{\epsilon \tau_0}{\gamma b} \leq \epsilon f(S^*)
\end{align*}

Therefore, Lemma \ref{lemma:r_2} mainly follows from case (2). \textbf{With {case (2)}, our proof is inspired from \cite{bian2017guarantees} and \cite{friedrich2019greedy}. We write down the detail of our proof for the sake of completeness.}

Denote $q$ as a minimum number in $0,...,t$ that there exists $i \in [k]$ in which $|S_q \cap V_i| = b_i$. Denote $i \in [k]$ as the index that $|S_q \cap V_i| = b_i$. For simplicity, for each $j=1,...,t$, denote $\rho_j = \Delta_{e_{j-1}} f(S_{j-1})$.  

\begin{claim} \label{claim:fast_lp_con}
For each $j=0,...,q-1$:
\begin{align*}
    f(S^*) \leq \sum_{r \leq j: e_r \in S^*} \rho_r + \alpha \sum_{r \leq j: e_r \not\in S^*} \rho_r + \frac{|S^* \setminus S_j|}{\gamma\sigma} \rho_{j+1}
\end{align*}
\end{claim}

\begin{proof}
Consider for each $j=0,...,q-1$, we have:
\begin{align}
    f(S^*) &= f(S^* \cup S_j) - \sum_{r \leq j} \Delta_{e_r} f(S^* \cup S_r) = f(S^* \cup S_j) - \sum_{r \leq j: e_r \not\in S^*} \Delta_{e_r} f(S^* \cup S_r) \\
    & \leq f(S^* \cup S_j) - (1-\alpha) \sum_{r \leq j: e_r \not\in S^*} \rho_{r+1} \label{equ:fast_con_1}
\end{align}
On the other hand, 
\begin{align}
    f(S^* \cup S_j) - f(S_j) &= \sum_{e^*_l \in S^*} \Delta_{e^*_l} f(S \cup \{e^*_1,...,e^*_{l-1}\}) \leq \frac{1}{\gamma\sigma} \sum_{e^*_l \in S^*} \Delta_{e^*_l} f(S_j) \leq \frac{|S^* \setminus S_j|}{\gamma\sigma} \rho_{j+1}  \label{equ:fast_con_2}
\end{align}
The claim follows by combining Equ. (\ref{equ:fast_con_1}), (\ref{equ:fast_con_2}) and the fact that $f(S_j) = \sum_{r \leq j} \rho_r$.
\end{proof}

For each $j=1,...,t$, denote $x_j = \frac{\rho_j}{f(S^*)}$; and $\beta_j = \alpha$ if $e_j \not\in S^*$, $1$ otherwise. Since $\frac{f(S)}{f(S^*)} = \sum_{j \leq t} x_j \geq \sum_{j \leq q} x_q$, by Claim \ref{claim:fast_lp_con}, $\frac{f(S)}{f(S^*)}$ is at least the value of the optimal solution of the following linear programming.
\begin{align}
    & \mbox{min} \quad \sum_{j=1}^q x_j  & \\
    & \mbox{s.t} \quad \sum_{r=1}^{j-1} \beta_r x_r + \frac{|S^* \setminus S_{j-1}|}{\gamma\sigma} x_j \geq 1 & \forall j=1,...,q \label{equ:con1_lp} \\
    & \quad \quad x_j \geq 0 & \forall j=1,...,q \label{equ:fast_lp_con_2}
\end{align}
Let's call this linear programming \textit{GrLP}. We have the following claim.
\begin{claim} \label{claim:fast_lp}
If $x^*_1,...,x^*_q$ is optimal solution of GrLP, then 
\begin{align*}
    \sum_{j=1}^q x^*_j \geq \frac{1}{\alpha} \Big[ 1 - \Big( 1 - \frac{\alpha\gamma\sigma}{b} \Big)^{q} \Big]
\end{align*}
\end{claim}

Claim \ref{claim:fast_lp} also concludes the proof for Lemma \ref{lemma:r_2}. Before proving Claim \ref{claim:fast_lp}, we have the following lemma, which is critical to obtain the optimal solution of GrLP.

\begin{lemma} \label{lemma:fast_lp_r}
For each $r=1,...,q$, if $e_r \in S^*$ then $x^*_r \leq x^*_{r+1}$.
\end{lemma}
\begin{proof}
We use contradiction: Assume $x^*_r > x^*_{r+1}$, then we can construct a feasible solution $y_1,...,y_q$ to GrLP such that $\sum_{r=1,...,q} y_r < \sum_{r=1,...,q} x^*_r$, thus contradicts to the fact that $x^*_1,....,x^*_q$ is the optimal one. 

Denote $p_t = |S^* \setminus S_t|$ for all $t=0,...,q$. It is trivial that $p_0,...,p_q$ is monotone non-increasing. Since $e_r \in S^*$, $p_{r-1} = p_r + 1$.

Let $\eta = \frac{(p_{r-1} - \sigma \gamma) x^*_r - p_r x^*_{r+1}}{p_{r-1}} \geq \frac{p_{r}}{p_{r-1}} (x^*_r - x^*_{r+1}) \geq 0$. The construction of $y_1,...,y_q$ is as follows:
\begin{itemize}
    \item $y_{r^\prime} = x^*_{r^\prime}$ for all $r^\prime < r$.
    \item $y_r = x^*_r - \eta$
    \item $y_{r^\prime} = x^*_{r^\prime} + \eta_{r^\prime}$ for all $r^\prime > r$, where
    \begin{itemize}
        \item $\eta_{r+1} = \eta \times \frac{\gamma\sigma}{p_r}$
        \item $\eta_{r + 1 + u} = \eta_{r + u} \times \frac{p_r - u + 1 - \gamma\sigma}{p_r - u}$
    \end{itemize}
\end{itemize}

\begin{claim}
$y_1,...,y_q$ satisfy the constraint (\ref{equ:fast_lp_con_2}) of GrLP.
\end{claim}

\begin{proof}
It is trivial that $y_{r^\prime} \geq x_{r^\prime} \geq 0$ for all $r^\prime \neq r$. On the other hand:
\begin{align*}
    y_r & = x^*_r - \eta = x^*_r - \frac{(p_{r-1} - \sigma \gamma) x^*_r - p_r x^*_{r+1}}{p_{r-1}} = \frac{\sigma\gamma x^*_r + p_r x^*_{r+1}}{p_{r-1}} \geq 0
\end{align*}
which completes the proof.
\end{proof}

\begin{claim}
$y_1,...,y_q$ satisfy the constraint (\ref{equ:con1_lp}) of GrLP.
\end{claim}

\begin{proof}
As $y_{r^\prime} = x^*_{r^\prime}$ for all $r^\prime < r$, $y_1,...,y_q$ satisfy the constraints of Equ. (\ref{equ:con1_lp}) for all $t < r$. In the following parts, we first prove the constraint (\ref{equ:con1_lp}) at $t=r$ is still satisfied by $y_1,...,y_q$. 

From the constraint (\ref{equ:con1_lp}) at $t=r+1$, we have that:
\begin{align}
    \sum_{a < r: e_a \in S^*} x^*_a + \alpha \sum_{a < r: e_a \not\in S^*} x^*_a + x^*_r + \frac{p_r}{\gamma\sigma} x^*_{r+1} \geq 1 \label{equ:con1_t_r+1}
\end{align}

Take the different between left-hand side (l.h.s) of the constraint (\ref{equ:con1_lp}) at $t=r$ with $y_1,...,y_q$ and l.h.s of (\ref{equ:con1_t_r+1}), we have
\begin{align*}
    & \Big[ \sum_{a < r: e_a \in S^*} y_a + \alpha \sum_{a < r: e_a \not\in S^*} y_a +  \frac{p_{r-1}}{\gamma\sigma} y_r \Big]  - \Big[ \sum_{a < r: e_a \in S^*} x^*_a + \alpha \sum_{a < r: e_a \not\in S^*} x^*_a + x^*_r + \frac{p_r}{\gamma\sigma} x^*_{r+1} \Big] \\
    & \quad = \frac{p_{r-1}}{\gamma\sigma} \Big( x^*_r - \frac{(p_{r-1} - \sigma \gamma) x^*_r - p_r x^*_{r+1}}{p_{r-1}} \Big) - x^*_r - \frac{p_r}{\gamma\sigma} x^*_{r+1} = 0
\end{align*}
Thus the constraint (\ref{equ:con1_lp}) at $t=r$ is still satisfied with $y_1,...,y_q$.

Next, we would need to prove that $y_1,...,y_q$ satisfy  the constraint (\ref{equ:con1_lp}) at $t > r$. Denote:
\begin{align*}
    \delta_{r+u} & = \Big(\mbox{l.h.s of (\ref{equ:con1_lp}) with $y_1,...,y_q$ at $t=r+u$}\Big) - \Big(\mbox{l.h.s of (\ref{equ:con1_lp}) with $x^*_1,...,x^*_q$ at $t=r+u$}\Big)
\end{align*}
To prove $y_1,...,y_q$ satisfy  the constraint (\ref{equ:con1_lp}) at $t > r$, we show that $\delta_{r+u}$ is monotone non-decreasing for all $u \geq 1$ and $\delta_{r+1} = 0$. 

Let's consider the constraint (\ref{equ:con1_lp}) at $t=r+1$ with $y_1,...,y_q$, we have:
\begin{align*}
    &\sum_{a < r: e_a \in S^*} y_a + \alpha \sum_{a < r: e_a \not\in S^*} y_a + y_r + \frac{p_r}{\gamma\sigma} y_{r+1} \\
    & \quad = \sum_{a < r: e_a \in S^*} x^*_a + \alpha \sum_{a < r: e_a \not\in S^*} x^*_a + x^*_r - \eta + \frac{p_r}{\gamma\sigma} ( x^*_{r+1} + \eta \frac{\gamma\sigma}{p_r} ) \\
    & \quad = \mbox{l.h.s of the constraint (\ref{equ:con1_lp}) at $t=r+1$ with $x^*_1,...,x^*_q$} \\
    & \quad \geq 1
\end{align*}
Thus, $\delta_{r+1} = 0$ and the constraint (\ref{equ:con1_lp}) at $t=r+1$ is still satisfied with $y_1,...,y_q$.

Next, we prove $\delta_{r+u+1} \geq \delta_{r+u}$ for all $u \geq 1$. The l.h.s of (\ref{equ:con1_lp}) with $y_1,...,y_q$ at $t=r+u$ is as follows:
\begin{align*}
    \sum_{a < r+u: e_a \in S^*} y_a + \alpha \sum_{a < r+u: e_a \not\in S^*} y_a + \frac{p_{r+u-1} }{\gamma\sigma} y_{r+u}
\end{align*}

In $t=r+u+1$, there will be 2 cases: (1) $e_{r+u} \in S^*$; and (2) $ e_{r+u} \not\in S^*$. 

With case (1), $p_{r+u} = p_{r+u-1} - 1$; the l.h.s of the constraint at $t= r+u+1$ becomes:
\begin{align*}
    \sum_{a < r+u: e_a \in S^*} y_a + \alpha \sum_{a < r+u: e_a \not\in S^*} y_a + y_{r+u} +  \frac{p_{r+u-1} - 1}{\gamma\sigma} y_{r+u+1}
\end{align*}
As $p_{r+u-1} \geq p_r - u + 1$, we have: 
\begin{align*}
    \delta_{r+u+1} - \delta_{r+u} & = \frac{p_{r+u-1} - 1}{\gamma\sigma} \eta_{r+u+1} - \frac{p_{r+u-1} - \gamma\sigma}{\gamma\sigma} \eta_{r+u} \\
    & = \Big[ (p_{r+u-1} - 1) \times \frac{p_r - u + 1 - \gamma\sigma}{p_r - u} -  (p_{r+u-1} - \gamma\sigma) \Big] \frac{\eta_{r+u}}{\gamma\sigma} \\
    & \geq \Big[ (p_{r+u-1} - 1) \times \frac{p_{r+u-1} - \gamma\sigma}{p_{r+u-1} - 1} -  (p_{r+u-1} - \gamma\sigma) \Big] \frac{\eta_{r+u}}{\gamma\sigma} \\
    & \geq 0
\end{align*}

With case (2), $p_{r+u} = p_{r+u-1}$; the l.h.s of the constraint at $t= r+u+1$ becomes:
\begin{align*}
    \sum_{a < r+u: e_a \in S^*} y_a + \alpha \sum_{a < r+u: e_a \not\in S^*} y_a + \alpha y_{r+u} +  \frac{p_{r+u-1}}{\gamma\sigma} y_{r+u+1}
\end{align*}
As $\eta_{r+u+1} \geq \eta_{r+u}$ , we have:
\begin{align*}
    \delta_{r+u+1} - \delta_{r+u} & = \frac{p_{r+u-1}}{\gamma\sigma} \eta_{r+u+1} - \frac{p_{r+u-1} - \alpha\gamma\sigma}{\gamma\sigma} \eta_{r+u} \\
    & \geq \frac{p_{r+u-1}}{\gamma\sigma} (\eta_{r+u+1} - \eta_{r+u}) \geq 0
\end{align*}
Therefore, $y_1,...,y_q$ satisfy the constraint (\ref{equ:con1_lp}).
\end{proof}

\begin{claim} \label{claim:fast_lp_y}
$\sum_{a=1}^q y_a \leq \sum_{a=1}^q x^*_a$
\end{claim}

\begin{proof}
We have $\sum_{a=1}^1 y_a = \sum_{a=1}^q x^*_a - \eta + \sum_{a=r+1}^q \eta_a$, where
\begin{align*}
    - \eta + \sum_{a=r+1}^q \eta_a = \eta \Big[ -1 + \frac{\gamma\sigma}{p_r} + \sum_{a=r+2}^q \frac{\gamma\sigma}{p_r} \prod_{l=1}^{a-r-1} \frac{p_r - l + 1 - \gamma\sigma}{p_r - l}  \Big]
\end{align*}
On the other hand,
\begin{align*}
    & -1 + \frac{\gamma\sigma}{p_r} + \sum_{a=r+2}^q \frac{\gamma\sigma}{p_r} \prod_{l=1}^{a-r-1} \frac{p_r - l + 1 - \gamma\sigma}{p_r - l} \\
    & = -1 + \frac{\gamma\sigma}{p_r} + \frac{\gamma\sigma}{p_r} \frac{p_r - \gamma\sigma}{p_r - 1} + ... + \frac{\gamma\sigma}{p_r} \frac{p_r - \gamma\sigma}{p_r - 1} ... \frac{p_r - q + r + 1 - \gamma\sigma}{p_r - q + r} \\ 
    & = - \frac{(p_r - \gamma\sigma)(p_r - 1 - \gamma\sigma) ... (p_r - q + r + 1 - \gamma\sigma)}{p_r(p_r - 1) ... (p_r - q + r)} < 0
\end{align*}
\end{proof}

Claim \ref{claim:fast_lp_y} means that not only $y_1,...,y_q$ is feasible to GrLP but also has better objective value than $x_1^*,...,x_q^*$, which violates the assumption that $x^*_1,...,x^*_q$ is the optimal solution. That completes the proof for Lemma \ref{lemma:fast_lp_r}. 
\end{proof}

In the next part, we will bound the value of $\sum_{j=1}^q x_j^*$. Denote $LP(\{a_1,...,a_l\})$ as the optimal result of GrLP if $\{a_1,...,a_l\} = \{r \mid e_r \in S^*\}$. For simplicity, we also refer $LP(\{a_1,...,a_l\})$ as GrLP if $\{a_1,...,a_l\} = \{r \mid e_r \in S^*\}$.

We will prove that $LP(\{a_1,...,a_l\}) \geq LP(\emptyset)$ by the following lemmas.

\begin{lemma} \label{lemma:fast_lp_right}
If there exists $s \leq l$, such that $a_s + 1 \not\in \{a_1,...,a_l\}$, then 
\begin{align*}
    LP(\{a_1,...,a_{s-1},a_s,a_{s+1},...,a_l\}) \geq LP(\{a_1,...,a_{s-1},a_s + 1,a_{s+1},...,a_l\})
\end{align*}
\end{lemma}

\begin{proof}
For simplicity, assume $x_1^*,...,x_q^*$ as the optimal solution of $LP(\{a_1,...,a_{s-1},a_s,a_{s+1},...,a_l\})$. We gonna prove $x_1^*,...,x_q^*$ is feasible to $LP(\{a_1,...,a_{s-1},a_s + 1,a_{s+1},...,a_l\})$.

It is trivial that $x_1^*,...,x_q^*$ satisfy the constraint (\ref{equ:con1_lp}) at $t=1,...,a_s$ of $LP(\{a_1,...,a_{s-1},a_s + 1,a_{s+1},...,a_l\})$.

From constraint (\ref{equ:con1_lp}) at $t=a_s+1$ of $LP(\{a_1,...,a_{s-1},a_s,a_{s+1},...,a_l\})$ we have:
\begin{align}
    \sum_{r < a_s: e_r \in S^*} x^*_r + \alpha \sum_{r < a_s: e_r \not\in S^*} x^*_r + x^*_{a_s} +  \frac{p_{a_s}}{\gamma\sigma} x^*_{a_s+1} \geq 1 \label{equ:lp1}
\end{align}
Plugging $x_1^*,...,x_q^*$ to the l.h.s of constraint (\ref{equ:con1_lp}) at $t=a_s+1$ of $ LP(\{a_1,...,a_{s-1},a_s + 1,a_{s+1},...,a_l\})$, we have:
\begin{align}
    \sum_{r < a_s: e_r \in S^*} x^*_r + \alpha \sum_{r < a_s: e_r \not\in S^*} x^*_r + \alpha x^*_{a_s} +  \frac{p_{a_s} + 1}{\gamma\sigma} x^*_{a_s+1} \label{equ:lp2}
\end{align}
The difference between (\ref{equ:lp2}) and the l.h.s of (\ref{equ:lp1})  is $\frac{1}{\gamma\sigma} x^*_{a_s + 1} - (1-\alpha) x_{a_s}^* \geq 0$, which means $x_1^*,...,x_q^*$ satisfy constraint (\ref{equ:con1_lp}) at $t=a_s+1$ of $ LP(\{a_1,...,a_{s-1},a_s + 1,a_{s+1},...,a_l\})$.

At $t > a_s + 1$, $x_1^*,...,x_q^*$ trivially satisfy since the difference between l.h.s of constraint (\ref{equ:con1_lp}) of $ LP(\{a_1,...,a_{s-1},a_s + 1,a_{s+1},...,a_l\})$ and the l.h.s of the one of $ LP(\{a_1,...,a_{s-1},a_s,a_{s+1},...,a_l\})$ is always $(1-\alpha)(x^*_{a_s + 1} - x^*_{a_s}) \geq 0$

Therefore, $x_1^*,...,x_q^*$ is feasible to $LP(\{a_1,...,a_{s-1},a_s + 1,a_{s+1},...,a_l\})$, which also means $LP(\{a_1,...,a_{s-1},a_s,a_{s+1},...,a_l\}) \geq LP(\{a_1,...,a_{s-1},a_s + 1,a_{s+1},...,a_l\})$.
\end{proof}



Lemma \ref{lemma:fast_lp_right} means that: as long as there exists $s$ that $a_s + 1 < a_{s+1}$, we keep increase the index of $a_s$ by 1 to obtain a new GrLP with lower optimal result. In the end, we have $LP(\{a_1,...,a_l\}) \geq LP(\{q-l+1,...,q\})$.

\begin{lemma} \label{lemma:fast_lp_remove}
$LP(\{q-l+1,...,q\}) \geq LP(\{q-l+2,...,q\})$
\end{lemma}
\begin{proof}
Denote $y_1^*,...,y_q^*$ as the optimal solution of $LP(\{q-l+1,...,q\})$. We need to prove $y_1^*,...,y_q^*$ is feasible to $LP(\{q-l+2,...,q\})$. 

It is trivial that $y_1^*,...,y_q^*$ satisfy the constraints (\ref{equ:con1_lp}) at $t=1,...,q - l + 1$ of $LP(\{q-l+2,...,q\})$.

At $t=q- l +2,...,q$, with $LP(\{q-l+1, q-l+2,...,q\})$ we have:
\begin{align}
    \alpha \sum_{r < q-l+1} y^*_r + y^*_{q-l+1} + \sum_{r=q-l+2}^{t}  y^*_r +  \frac{B - t+q-l+1}{\gamma\sigma} y^*_{t} \geq 1 \label{equ:lp3}
\end{align}
Plugging $y_1^*,...,y_q^*$ to the l.h.s of this constraint but with $ LP(\{q-l+2,...,q\})$, we have:
\begin{align}
    \alpha \sum_{r < q-l+1} y^*_r + \alpha y^*_{q-l+1} +  \sum_{r=q-l+2}^{t}  y^*_r +  \frac{B-t+q-l+2}{\gamma\sigma} y^*_{t} \label{equ:lp4}
\end{align}
The difference between (\ref{equ:lp4}) and the l.h.s of (\ref{equ:lp3}), therefore, is $\frac{1}{\gamma\sigma} y^*_{t} - (1-\alpha) y^*_{q-l+1} \geq 0$ since $y^*_{q-l+1} \leq y^*_{q-l+2} \leq ... \leq y^*_{q}$ (Lemma \ref{lemma:fast_lp_r}).

Therefore, $y_1^*,...,y_q^*$ is feasible to $LP(\{q-l+2,...,q\})$, which means $LP(\{q-l+1,...,q\}) \geq LP(\{q-l+2,...,q\})$.
\end{proof}


Therefore, from Lemma \ref{lemma:fast_lp_remove}, we have
\begin{align*}
    LP(\{q-l+1,...,q\}) \geq LP(\{q-l+2,...,q\}) \geq ... \geq LP(\emptyset)
\end{align*}

GrLP with $LP(\emptyset)$ is formulated as follows:
\begin{align*}
    \mbox{min} & \quad \sum_{r=1,...,q} x_r  &\\
    \mbox{s.t} & \quad \alpha \sum_{r < t} x_r + \frac{b}{\gamma\sigma} x_t \geq 1 & \forall t=1,...,q  \\
    & \quad x_r \geq 0 & \forall r=1,...,q 
\end{align*}

Trivially, $LP(\emptyset)$ reaches optimum at $x_r = \frac{\gamma\sigma}{b} \Big( 1-\frac{\alpha\gamma\sigma}{b} \Big)^{r-1}$ for $r=1,...,q$, where
\begin{align*}
    LP(\emptyset) = \sum_{r=1}^q \frac{\gamma\sigma}{b} \Big( 1-\frac{\alpha\gamma\sigma}{b} \Big)^{r-1} = \frac{1}{\alpha} \Big[ 1 - \Big( 1 - \frac{\alpha\gamma\sigma}{b} \Big)^q \Big]
\end{align*}
which concludes the proof for claim \ref{claim:fast_lp}.

The query complexity of \gr can be trivially inferred. With \fast, the value of $\tau$ reduces by a factor of $1-\epsilon$ after each \textbf{while} loop (line \ref{line:fast_loop} Alg. \ref{alg:fast}). Therefore, \fast runs at most $O(\frac{1}{\epsilon} \ln b)$ loops, where each loop queries $f$ at most $O(n)$ times. Therefore, \fast's query complexity is $O(\frac{n}{\epsilon} \ln b)$, which completes the proof.

\end{proof}

\section{Omitted Proofs of Experiment Section} \label{apd:exp}

\subsection{Proof of Lemma \ref{lemma:boost_bound}} 
A graph realization $g$ of $G$ is defined as a subgraph of $G$ where edges' state were determined. Denote $R_g(I)$ as a number of reachable nodes from $I$ in $g$. Given a boosted set $H$, the probability $g$ is formed is given by
\begin{align*}
    \rho_H(g) = \prod_{e \in g} p_H(e) \prod_{e \not\in g} \big(1-p_H(e) \big)
\end{align*}
where $p_H(e) = p_e^0$ if $e=(u,v)$ has $v \not\in H$; otherwise $p_H(e) = \rho_e^1$.

Let $\mathcal{G}$ be a collection of all graph realization of $G$. We have:
\begin{align*}
    f(H) = \sum_{g \in \mathcal{G}} \rho_H(g) \times R_g(I)
\end{align*}

Given sets $W, H$ that $H \subset W$ and $|(W \setminus H) \cap V_i| \leq b_i$ for all $i \in [k]$, $u \not\in W$ and a graph realization $g$. For each $v \in W$, any edge in form $e=(w,v)$ satisfies $p_H(e) = p_{H \cup \{u\}}(e)$ and $p_W(e) = p_{W \cup \{u\}}(e)$. On the other hand, any edge in form $e=(w,u)$ will have weights that  $p_H(e) = p_W(e) = p_e^0$ and $p_{H \cup \{u\}}(e)= p_{W \cup \{u\}}(e) = p_e^1$. Therefore, we always have $\frac{p_{W \cup \{u\}}(e)}{p_{H \cup \{u\}}(e)} = \frac{p_{H}(e)}{p_{W}(e)}$ and $\frac{1 -p_{W \cup \{u\}}(e)}{1 -p_{H \cup \{u\}}(e)} = \frac{1- p_{H}(e)}{1- p_{W}(e)}$ for all $e \in g$, which means $\frac{\rho_W(g)}{\rho_H(g)} = \frac{\rho_{W \cup \{u\}}(g)}{\rho_{H \cup \{u\}}(g)}$. 

On the other hand:
\begin{align*}
    \frac{\rho_W(g)}{\rho_H(g)} = \prod_{e=(w,v) \in g; v\in W\setminus H} \frac{\rho_e^1}{\rho_e^0} \prod_{e=(w,v) \not\in g; v\in W\setminus H} \frac{1-\rho_e^1}{1-\rho_e^0}
\end{align*}
Since $|W \setminus H| \leq b$, we have
\begin{align}
    \min_{E^\prime \subseteq E: |E^\prime| \leq b\Delta} \prod_{e \in E^\prime} \frac{1 - p_{e}^1}{1 - p_{e}^0} \leq \frac{\rho_W(g)}{\rho_H(g)} &\leq \max_{E^\prime \subseteq E: |E^\prime| \leq b\Delta} \prod_{e \in E^\prime} \frac{p_e^1}{p_e^0} \label{equ:boost_gamma}
\end{align}

We have:
\begin{align}
    f(W \cup \{u\}) - f(W) & = \sum_{g \in \mathcal{G}} \big( \rho_{W \cup \{u\}}(g) - \rho_W(g) \big) \times R_g(S) \\
    &  = \sum_{g \in \mathcal{G}} \Big[ \frac{\rho_{W \cup \{u\}}(g)}{\rho_{H \cup \{u\}}(g)} \rho_{H \cup \{u\}}(g)  - \frac{\rho_W(g)}{\rho_H(g)} \rho_H(g) \Big] R_g(S) \\
    &  = \sum_{g \in \mathcal{G}} \frac{\rho_W(g)}{\rho_H(g)} \big( \rho_{H \cup \{u\}}(g) - \rho_{H}(g)\big) \times R_g(S)\label{equ:boost_gain}
\end{align}
The lemma follows by combining Equ. (\ref{equ:boost_gamma}) and (\ref{equ:boost_gain}).







\subsection{Proof of Lemma \ref{lemma:video_bound}} 
For any set $S$, denote $A_S = I + X_S$. Given a set $T$, $S \subset T$ and $e \not\in T$, we have:
\begin{align*}
    f(S \cup \{e\}) - f(S) &= \prod_{i=1}^{|S \cup \{e\}|} \lambda_i(A_{S \cup \{e\}}) - \prod_{i=1}^{|S|} \lambda_i(A_S)\\
    & = \lambda_{|S \cup \{e\}|}(A_{S \cup \{e\}}) \prod_{i=1}^{|S|} \lambda_i(A_{S \cup \{e\}}) - \prod_{i=1}^{|S|} \lambda_i(A_S) \\
    &  \geq \Big( \lambda_{|S \cup \{e\}|}(A_{S \cup \{e\}}) - 1 \Big) \prod_{i=1}^{|S|} \lambda_i(A_S)
\end{align*}
where the last inequality comes from Cauchy interlacing inequality. On the other hand,
\begin{align*}
    f(T \cup \{e\}) - f(T) & = \prod_{i=1}^{|T \cup \{e\}|} \lambda_i(A_{T \cup \{e\}}) - \prod_{i=1}^{|T|} \lambda_i(A_T) \\
    & = \lambda_1(A_{T \cup \{e\}}) \prod_{i=2}^{|T \cup \{e\}|} \lambda_i(A_{T \cup \{e\}}) - \prod_{i=1}^{|T|} \lambda_i(A_T) \\
    &  \leq \Big( \lambda_1(A_{T \cup \{e\}}) - 1 \Big) \prod_{i=1}^{|T|} \lambda_i(A_T) \\
    & = \Big( \lambda_1(A_{T \cup \{e\}}) - 1 \Big) \prod_{i=|T| - |S| + 1}^{|T|} \lambda_i(A_T) \prod_{i=1}^{|T|-|S|} \lambda_i (A_T) \\
    &  \leq \Big( \lambda_1(A_{T \cup \{e\}}) - 1 \Big) \prod_{i=1}^{|S|} \lambda_i(A_S) \prod_{i=1}^{|T|-|S|} \lambda_i (A_T) 
\end{align*}

Therefore:
\begin{align*}
    \frac{f(T \cup \{e\}) - f(T)}{f(S \cup \{e\}) - f(S)} & \leq \frac{\Big( \lambda_1(A_{T \cup \{e\}}) - 1 \Big) \prod_{i=1}^{|S|} \lambda_i(A_S) \prod_{i=1}^{|T|-|S|} \lambda_i (A_T)}{\Big( \lambda_{|S \cup \{e\}|}(A_{S \cup \{e\}}) - 1 \Big) \prod_{i=1}^{|S|} \lambda_i(A_S)} \\
    & \leq \frac{ \lambda_1(A) - 1 }{  \lambda_n(A) - 1 } \prod_{i=1}^{b} \lambda_i(A)
\end{align*}
which completes the proof.

%% file: ijcai22-full.bbl
\begin{thebibliography}{}

\bibitem[\protect\citeauthoryear{Badanidiyuru and
  Vondr{\'a}k}{2014}]{badanidiyuru2014fast}
Ashwinkumar Badanidiyuru and Jan Vondr{\'a}k.
\newblock Fast algorithms for maximizing submodular functions.
\newblock In {\em Proceedings of the twenty-fifth annual ACM-SIAM symposium on
  Discrete algorithms}, pages 1497--1514. SIAM, 2014.

\bibitem[\protect\citeauthoryear{Bian \bgroup \em et al.\egroup
  }{2017}]{bian2017guarantees}
Andrew~An Bian, Joachim~M Buhmann, Andreas Krause, and Sebastian Tschiatschek.
\newblock Guarantees for greedy maximization of non-submodular functions with
  applications.
\newblock In {\em International conference on machine learning}, pages
  498--507. PMLR, 2017.

\bibitem[\protect\citeauthoryear{Bogunovic \bgroup \em et al.\egroup
  }{2017}]{bogunovic2017robust}
Ilija Bogunovic, Slobodan Mitrovi{\'c}, Jonathan Scarlett, and Volkan Cevher.
\newblock Robust submodular maximization: A non-uniform partitioning approach.
\newblock In {\em Proceedings of the 34th International Conference on Machine
  Learning-Volume 70}, pages 508--516. JMLR. org, 2017.

\bibitem[\protect\citeauthoryear{Bogunovic \bgroup \em et al.\egroup
  }{2018}]{bogunovic2018robust}
Ilija Bogunovic, Junyao Zhao, and Volkan Cevher.
\newblock Robust maximization of non-submodular objectives.
\newblock In {\em International Conference on Artificial Intelligence and
  Statistics}, pages 890--899, 2018.

\bibitem[\protect\citeauthoryear{Buchbinder \bgroup \em et al.\egroup
  }{2014}]{buchbinder2014submodular}
Niv Buchbinder, Moran Feldman, Joseph Naor, and Roy Schwartz.
\newblock Submodular maximization with cardinality constraints.
\newblock In {\em Proceedings of the twenty-fifth annual ACM-SIAM symposium on
  Discrete algorithms}, pages 1433--1452. SIAM, 2014.

\bibitem[\protect\citeauthoryear{Buchbinder \bgroup \em et al.\egroup
  }{2019}]{buchbinder2019deterministic}
Niv Buchbinder, Moran Feldman, and Mohit Garg.
\newblock Deterministic ($1/2$+ $\varepsilon$)-approximation for submodular
  maximization over a matroid.
\newblock In {\em Proceedings of the Thirtieth Annual ACM-SIAM Symposium on
  Discrete Algorithms}, pages 241--254. SIAM, 2019.

\bibitem[\protect\citeauthoryear{Calinescu \bgroup \em et al.\egroup
  }{2011}]{calinescu2011maximizing}
Gruia Calinescu, Chandra Chekuri, Martin Pal, and Jan Vondr{\'a}k.
\newblock Maximizing a monotone submodular function subject to a matroid
  constraint.
\newblock {\em SIAM Journal on Computing}, 40(6):1740--1766, 2011.

\bibitem[\protect\citeauthoryear{Chen \bgroup \em et al.\egroup
  }{2018}]{chen2018weakly}
Lin Chen, Moran Feldman, and Amin Karbasi.
\newblock Weakly submodular maximization beyond cardinality constraints: Does
  randomization help greedy?
\newblock In {\em International Conference on Machine Learning}, pages
  804--813, 2018.

\bibitem[\protect\citeauthoryear{Conforti and
  Cornu{\'e}jols}{1984}]{conforti1984submodular}
Michele Conforti and G{\'e}rard Cornu{\'e}jols.
\newblock Submodular set functions, matroids and the greedy algorithm: tight
  worst-case bounds and some generalizations of the rado-edmonds theorem.
\newblock {\em Discrete applied mathematics}, 7(3):251--274, 1984.

\bibitem[\protect\citeauthoryear{Cornnejols \bgroup \em et al.\egroup
  }{1977}]{cornnejols1977location}
G~Cornnejols, M~Fisher, and G~Nemhauser.
\newblock Location of bank accounts of optimize float: An analytic study of
  exact and approximate algorithm.
\newblock {\em Management Science}, 23:789--810, 1977.

\bibitem[\protect\citeauthoryear{Das and Kempe}{2011}]{das2011submodular}
Abhimanyu Das and David Kempe.
\newblock Submodular meets spectral: greedy algorithms for subset selection,
  sparse approximation and dictionary selection.
\newblock In {\em Proceedings of the 28th International Conference on
  International Conference on Machine Learning}, pages 1057--1064, 2011.

\bibitem[\protect\citeauthoryear{Elenberg \bgroup \em et al.\egroup
  }{2017}]{elenberg2017streaming}
Ethan Elenberg, Alexandros~G Dimakis, Moran Feldman, and Amin Karbasi.
\newblock Streaming weak submodularity: Interpreting neural networks on the
  fly.
\newblock In {\em Advances in Neural Information Processing Systems}, pages
  4044--4054, 2017.

\bibitem[\protect\citeauthoryear{Friedrich \bgroup \em et al.\egroup
  }{2019}]{friedrich2019greedy}
Tobias Friedrich, Andreas G{\"o}bel, Frank Neumann, Francesco Quinzan, and Ralf
  Rothenberger.
\newblock Greedy maximization of functions with bounded curvature under
  partition matroid constraints.
\newblock In {\em Proceedings of the AAAI Conference on Artificial
  Intelligence}, volume~33, pages 2272--2279, 2019.

\bibitem[\protect\citeauthoryear{Gatmiry and
  Gomez-Rodriguez}{2018}]{gatmiry2018non}
Khashayar Gatmiry and Manuel Gomez-Rodriguez.
\newblock Non-submodular function maximization subject to a matroid constraint,
  with applications.
\newblock {\em arXiv preprint arXiv:1811.07863}, 2018.

\bibitem[\protect\citeauthoryear{Iyer \bgroup \em et al.\egroup
  }{2013}]{iyer2013curvature}
Rishabh~K Iyer, Stefanie Jegelka, and Jeff~A Bilmes.
\newblock Curvature and optimal algorithms for learning and minimizing
  submodular functions.
\newblock {\em Advances in Neural Information Processing Systems},
  26:2742--2750, 2013.

\bibitem[\protect\citeauthoryear{Kuhnle \bgroup \em et al.\egroup
  }{2018}]{kuhnle2018fast}
Alan Kuhnle, J~David Smith, Victoria Crawford, and My~Thai.
\newblock Fast maximization of non-submodular, monotonic functions on the
  integer lattice.
\newblock In {\em International Conference on Machine Learning}, pages
  2786--2795, 2018.

\bibitem[\protect\citeauthoryear{Lehmann \bgroup \em et al.\egroup
  }{2006}]{lehmann2006combinatorial}
Benny Lehmann, Daniel Lehmann, and Noam Nisan.
\newblock Combinatorial auctions with decreasing marginal utilities.
\newblock {\em Games and Economic Behavior}, 55(2):270--296, 2006.

\bibitem[\protect\citeauthoryear{Leskovec and Krevl}{2014}]{snapnets}
Jure Leskovec and Andrej Krevl.
\newblock {SNAP Datasets}: {Stanford} large network dataset collection.
\newblock \url{http://snap.stanford.edu/data}, June 2014.

\bibitem[\protect\citeauthoryear{Lin \bgroup \em et al.\egroup
  }{2017}]{lin2017boosting}
Yishi Lin, Wei Chen, and John~CS Lui.
\newblock Boosting information spread: An algorithmic approach.
\newblock In {\em 2017 IEEE 33rd International Conference on Data Engineering
  (ICDE)}, pages 883--894. IEEE, 2017.

\bibitem[\protect\citeauthoryear{Mirzasoleiman \bgroup \em et al.\egroup
  }{2016}]{mirzasoleiman2016fast}
Baharan Mirzasoleiman, Morteza Zadimoghaddam, and Amin Karbasi.
\newblock Fast distributed submodular cover: Public-private data summarization.
\newblock In {\em Advances in Neural Information Processing Systems}, pages
  3594--3602, 2016.

\bibitem[\protect\citeauthoryear{Qian \bgroup \em et al.\egroup
  }{2015}]{qian2015subset}
Chao Qian, Yang Yu, and Zhi-Hua Zhou.
\newblock Subset selection by pareto optimization.
\newblock In {\em Advances in Neural Information Processing Systems}, pages
  1774--1782, 2015.

\bibitem[\protect\citeauthoryear{Qian \bgroup \em et al.\egroup
  }{2018}]{qian2018multiset}
Chao Qian, Yibo Zhang, Ke~Tang, and Xin Yao.
\newblock On multiset selection with size constraints.
\newblock In {\em AAAI}, pages 1395--1402, 2018.

\end{thebibliography}
